\documentclass[aps,pr,twocolumn,amsmath,amssymb,longbibliography,noeprint,maxbibnames=2]{revtex4-1}

\usepackage[%
  colorlinks=true,
  urlcolor=blue,
  linkcolor=blue,
  citecolor=blue
]{hyperref}

\usepackage{graphicx}% Include figure files
\usepackage{dcolumn}% Align table columns on decimal point
\usepackage{bm}% bold math
\usepackage{ulem}
\usepackage{braket}
\usepackage[caption=false]{subfig}
\usepackage[compat=1.0.0]{tikz-feynman}
\graphicspath{ {./pics/} }
\begin{document}
%\preprint{AIP/123-QED}

\title{Demonstration of resonant inelastic X-ray scattering as a probe of exciton-phonon coupling}
\author{Andrey Geondzhian and Keith Gilmore}
\affiliation{European Synchrotron Radiation Facility, 71 avenue des Martyrs, 38043 Grenoble, France}
\date{\today}
\begin{abstract}
Resonant inelastic X-ray scattering (RIXS) is a promising technique for obtaining electron-phonon coupling constants.  However, the ability to extract these coupling constants throughout the Brillouin zone for crystalline materials remains limited.  To address this need, we developed a Green's function formalism to capture electron-phonon contributions to core-level spectroscopies without explicitly solving the full vibronic problem.  Our approach is based on the cumulant expansion of the Green's function combined with many-body theory calculated vibrational coupling constants.  The methodology is applied to X-ray photoemission spectroscopy, X-ray absorption spectroscopy (XAS), and RIXS.  In the case of the XAS and RIXS, we use a 2-particle exciton Green's function, which accounts implicitly for particle-hole interference effects that have previously proved difficult.  To demonstrate the methodology and gain a deeper understanding of the experimental technique, we apply our formalism to small molecules, for which unambiguous experimental data exist.  This comparison reveals that the vibronic coupling constant probed by RIXS is in fact related to exciton-phonon coupling rather than electron-phonon coupling, challenging the conventional interpretation of the experiment.
\end{abstract}

%\pacs{}
%\keywords{}
\maketitle

\section{introduction}

% Paragraph on importance of electron-phonon coupling
Electron-phonon coupling (EPC) is an inescapable aspect of condensed matter systems that has a pronounced influence on many quantities of interest.  Interactions between electrons and phonons induce structural and magnetic transitions, and contribute to the temperature dependence of charge and spin transport \cite{Migdal1958,Eliashberg1960,Millis1996,Grimaldi2006,Pascual-Gutierrez2009,Giustino2017a}.  Notably, exchange of phonons produces the attractive interaction between electrons that binds Cooper pairs in conventional BCS superconductors \cite{Bardeen1957}.  As observed in photoemission and inverse photoemission, phonons renormalize quasiparticle energies and spectral weights, and bestow them with lifetimes \cite{Hengsberger1999,Giustino2010,Bancroft2009}.  Phonons can accompany optical transitions, modifying both transition energies and probabilities \cite{Perebeinos2005}, effects that make particularly significant contributions for indirect-gap systems and organic molecules.

% Paragraph on measuring electron-phonon coupling
Several experimental techniques are able to measure electron-phonon coupling strengths to varying degrees of efficacy.  These include inelastic neutron scattering \cite{Pintschovius2005} and conventional inelastic X-ray scattering \cite{DA2003,Baron2007}, one electron addition or removal processes such as scanning tunneling spectroscopy \cite{Fischer2007,Schrieffer1962}, electron energy loss spectroscopy \cite{Tanaka2017}, angle resolved photoemission (ARPES) and inverse photoemission \cite{ipes1994,Zhang2017}, and techniques involving neutral excitations including Raman and optical spectroscopies \cite{Ferrari2007,Perebeinos2005}. All of these approaches have certain unsatisfactory limitations with respect to accurately quantifying the electron-phonon interaction strength throughout the full Brillouin zone.  Resonant inelastic X-ray scattering (RIXS) -- which can generate collective excitations by the perturbation present in the intermediate, resonantly core-excited state -- has emerged in the last decade as a new technique for accomplishing this objective \cite{Ament2010,Devereaux2016}.  RIXS holds certain advantages over these other techniques including element and orbital selectivity, sensitivity to small sample quantities, and momentum resolution.  Most importantly, it is commonly assumed that RIXS offers a {\it direct} probe of electron-phonon coupling \cite{Ament2011, Devereaux2016}, contrary to the indirect nature of some of the other measurements.  

With the renewed consideration of the possible role of electron-phonon coupling in Cooper pairing in unconventional superconductors \cite{LeTacon2014}, there is a desire to use RIXS to quantify momentum-dependent electron-phonon coupling strengths in these materials \cite{Devereaux2016}.  Within typical RIXS analysis \cite{Ament2011}, the coupling strength between the electronic system and the lattice is extracted as a fitting parameter and interpreted as the usual transport electron-phonon coupling constant.  This is based on the assumption that while the RIXS intermediate state contains both a core-hole and excited electron, the core-hole is fully screened and effectively inert with respect to producing phonons. This assertion has gone without rigorous investigation despite the importance of the subject.  This is due partly to the difficulty of applying first-principles methods to strongly correlated materials.  Therefore, to investigate the phonon generating process in a RIXS measurement and the role played by the core-hole we turn to a simpler system, namely acetone, for which high quality experimental RIXS data is available showing a series of well-resolved phonon features \cite{Schreck2016}.

In Sec.~\ref{sec:intro}, we use density functional theory (DFT) calculations to demonstrate quantitatively that the usual interpretation of the EPC parameter measured by RIXS as the transport EPC parameter cannot explain previously reported RIXS data for acetone.  We further show that assuming an EPC value calculated as an {\it exciton}-phonon coupling parameter yields a RIXS spectrum in close agreement with the experimental result.  This revised interpretation of the meaning of the coupling constant extracted from RIXS measurements as an exciton-phonon coupling value rather than an electron-phonon coupling parameter is fully consistent with the established understanding of Raman spectroscopy \cite{Compaan1972,Sceats1975,Trallero-Giner1989,Gillet2017}, which is essentially the optical analogue of RIXS. 

This observation that the electron-lattice interaction probed by RIXS differs from what is typically assumed provides further motivation for developing effective methods for calculating the phonon contribution to RIXS.  Such calculations typically work within a Hilbert space that explicitly contains the vibrational degrees of freedom.  This has been pursued, for example, by {\it ab initio} many-electron vibronic wavefunction techniques for small molecules \cite{Couto2017}, and through exact diagonalization of extended Hubbard models for strongly correlated materials \cite{Johnston2016, Devereaux2016}.  Both approaches suffer from rapid expansion of the Hilbert space making them poorly suited for scaling up to the general case of treating the full {\bf q}-dependence of several phonon modes that may contain multiple quanta of excitations in the RIXS final state.  This inability to perform a detailed analysis on a wide range of systems of interest has limited the use of RIXS to explore vibronic coupling in materials.

Our objective is to construct a practical prescription for including vibrational contributions to core-level spectra broadly, and RIXS in particular, that can be applied to general periodic systems.  We present an alternate route than those mentioned above in which explicit reference to vibrational modes is largely removed, but phonons are accounted for implicitly as giving contributions to the electronic Green's functions and associated spectral functions.  Within perturbation theory, to obtain satisfactory spectral functions that exhibit accurate phonon satellite structure it is necessary to go beyond the typical Dyson-Migdal Green's function approach.  To accomplish this, we employ the cumulant expansion for the Green's function, which has recently been applied to study vibronic coupling in valence-level photoemission \cite{Story2014,Nery2017}.

We begin this effort in Sec.~\ref{XPSsect} by considering vibrational sidebands in X-ray photoemission spectra (XPS).  We show how these phonon sidebands can be calculated from first-principles via a vibronically-dressed core-hole Green's function formalism.  These parameter free calculations are in excellent agreement with high resolution experimental data available for the Si 2$p$ XPS of SiH$_{4}$ and SiF$_{4}$ molecules.  In Sec.~\ref{xas} we generalize the 1-particle core-hole Green's function of XPS to a 2-particle exciton Green's function suitable for X-ray absorption (XAS) or emission (XES).  We perform frozen-phonon Bethe-Salpeter calculations of the absorption spectrum to construct a vibronic self-energy for the exciton.  The methodology is applied to the N-K edge XAS of N$_{2}$ and the O-K edge XAS of acetone and CO.  To evaluate the vibrational contribution to RIXS, we find it most effective to employ a mixed representation of the RIXS cross section in which explicit intermediate states are removed in favor of a Green's function, but final vibronic states are kept.  This is described in Sec.~\ref{RIXS} using the vibronically-dressed exciton Green's function and applied again to the O-K edge of acetone.  

Our application of the methodology developed here to small molecules is intended only to clearly demonstrate the validity of the technique.  The advantage of small molecules as test cases is that their electronic and vibrational structure is relatively simple, and experimental data showing well resolved vibrational features is available.  However, the methodology is meant to be applied to periodic systems.  While we envisage that predictive calculations would be based in density functional theory, any electronic structure technique that contains an explicit dependence on atomic positions could be used.

\section{\label{sec:intro} RIXS as a probe of Exciton-phonon coupling}
%S ~\ref{sec:intro}

% P on introduction to RIXS formalism
The resonant inelastic X-ray scattering cross-section is given formally by the Kramers-Heisenberg equation \cite{Kramers1925}
\begin{equation}
\label{KH-eq}
\sigma(\omega_i,\omega_{loss})=\sum_F \left | \sum_M\frac{\bra{\Psi_F}\Delta_2^+\ket{\Psi_M}\bra{\Psi_M}\Delta_1\ket{\Psi_I}}{\omega_i-(E_{M}-E_{I})+i\Gamma_M} \right |^2 $$
$$\times \delta(\omega_{loss}-(E_{F}-E_{I}))
\end{equation}
which includes a summation over many-body intermediate ($M$) and final ($F$) states, assuming the system begins in a particular initial state ($I$).  The many-body energies of these states are given by $E_S$ for $S \in \{I,M,F\}$, $\Gamma_M$ is the inverse lifetime of the intermediate state, $\omega_i$ is the incident photon energy, $\omega_{loss}$ the energy transfer, and $\Delta_j$ the photon operator for the incident ($j$=1) or scattered ($j$=2) photon.  To focus on the phonon contribution to the RIXS spectrum one can apply the Kramers-Heisenberg expression to an effective Hamiltonian with a first-order electron-phonon interaction 
~
\begin{equation}
\label{Frohlich}
H=\sum_{n{\bf k}} \epsilon_{n{\bf k}} c^{+}_{n{\bf k}} c_{n{\bf k}} + \sum_{\nu {\bf q}} \omega_{{\bf q}\nu} \left ( b^{+}_{{\bf q}\nu}b_{{\bf q}\nu} + 1/2 \right ) $$
$$ + \sum_{nn^{\prime}{\bf k}} \sum_{\nu{\bf q}} M_{nn^{\prime}}^{\nu}({\bf k},{\bf q}) c^{+}_{n^{\prime}{\bf k}+{\bf q}} c_{n{\bf k}} (b^{+}_{-{\bf q}\nu} + b_{{\bf q}\nu}) \, .
\end{equation}
$M_{nn^{\prime}}^{\nu}({\bf k},{\bf q})$ is the coupling constant between electrons indexed by band $n$ and wavevector ${\bf k}$ with energy $\epsilon_{n{\bf k}}$, created by $c_{n{\bf k}}^+$, and phonons of wavevector ${\bf q}$ and branch $\nu$ having frequency $\omega_{{\bf q}\nu}$, created by $b_{{\bf q}\nu}^+$.

It is common within RIXS analysis to reduce the problem to a Holstein Hamiltonian consisting of a single electronic state interacting with a single mode
~
\begin{equation}
\label{reducedHam}
H=\epsilon_{i} c^{+}_{i} c_{i} + \omega_{0} b^{+}_{i}b_{i} + M c^{+}_{i} c_{i} (b^{+}_{i} + b_{i}) .
\end{equation}
With the usual assumptions that there is no electronic interaction between orbitals and the vibrational mode does not scatter an electron between orbitals or sites (no recoil), that the ground- and excited-state vibrational potential energy surfaces are harmonic with equivalent curvature, and that the core-hole may be neglected, it becomes possible to diagonalize this Hamiltonian by a canonical transformation \cite{Lang1963,Mahan1990}.  Taking the low temperature limit that the system begins in the zero oscillator level, this brings the phonon contribution to the Kramers-Heisenberg RIXS cross section to the form
\begin{eqnarray}
\sigma(\omega_i,\omega_{loss}) &=& \sum_{n_f} \left | \sum_{n_m}\frac{B_{n^{\prime\prime}n^{\prime}}(g)B_{n_{m}0}(g)}{\omega_i-(g-n_{m})\omega_0+i\gamma_m} \right |^2 \nonumber \\
 && \times \, \delta(\omega_{loss}-n_{f}\omega_{0})
\label{AmentEq}
\end{eqnarray}
where $n^{\prime}={\rm min}(n_m,n_f)$, $n^{\prime\prime}={\rm max}(n_m,n_f)$, and $g=(M/\omega_0)^2$ is the dimensionless coupling strength.  The phonon contribution may then be evaluated analytically using the Franck-Condon factors
\begin{equation}
\label{FC-factor}
B_{mn}(g)=(-1)^m \sqrt{e^{-g}m!n!}\sum_{l=0}^{n} \frac{(-g)^{l}\sqrt{g}^{m-n}}{(n-l)!l!(m-n+l)!} \, .
\end{equation}
When the ground- and excited-state potential energy surfaces differ it is necessary to instead use a more general form of the Franck-Condon factors \cite{Chang2005}.

\begin{figure}
\center{\includegraphics[width=1\linewidth]{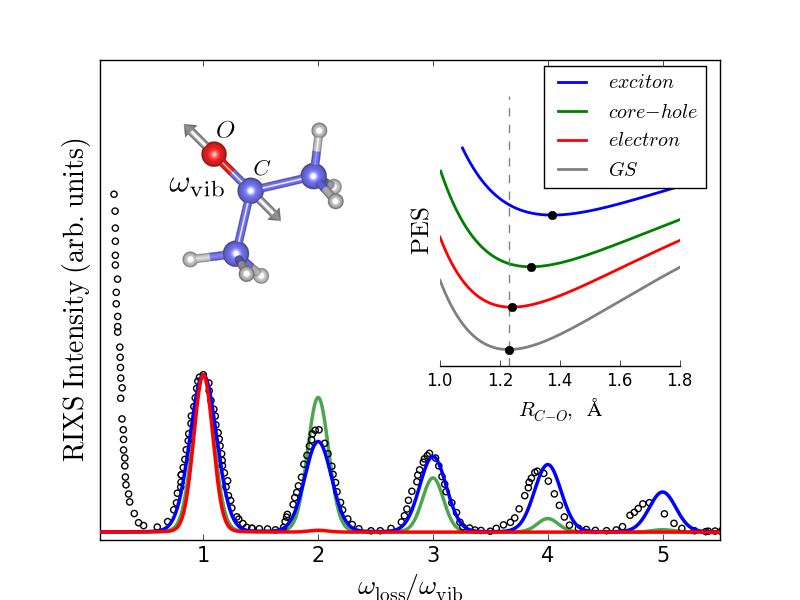}}	

\caption{RIXS spectrum at the O-K edge of acetone showing a progression of phonon excitations separated by the vibrational energy $\omega_{vib}$ = 210 meV.  Calculations resulting from an electron-phonon coupling constant assuming an intermediate state with an excited electron (red), an oxygen core-hole (green), and an exciton (blue) are compared to experimental results (open symbols) \cite{Schreck2016}. Calculated spectra are shown without the elastic contribution and are normalized to the intensity of the first phonon peak.  Inset: calculated potential energy surfaces (PES) for the three possible excited states along with the ground-state (GS) PES (gray).  The dashed vertical line indicates the ground-state equilibrium C=O bond length and the black dots are placed at the minima of each PES.  All PES curves are offset arbitrarily in energy.}
\label{modelRIXS}
\end{figure}

% Paragraph on Ament model
The above is the basis for the model presented by Ament {\it et al.~}\cite{Ament2011} that is commonly used to extract electron-phonon coupling values.  Equation \eqref{AmentEq} produces a RIXS intensity that consists of a series of peaks separated by the phonon energy with relative intensities that depend on the electron-phonon coupling strength (see Fig.~\ref{modelRIXS}).  Fitting experimental data with this series by tuning $g$ yields a value for the electron-phonon coupling constant. This and closely related approaches have been used to quantify the electron-phonon coupling strength in titanates and 1D cuprates \cite{Lee2013, Moser2015a, Fatale2016, Johnston2016a}, and have influenced the conceptual understanding of vibronic effects in RIXS \cite{Yavas2010, Ament2010}.

% P on calculation of coupling constants
Assuming the validity of the interpretation of $g$ in Eq.~\eqref{AmentEq} as the usual transport electron-phonon coupling strength, it should be possible, for an amenable system, to calculate the electron-phonon coupling constant {\it ab initio} and use the resulting value of $g$ as a fixed parameter in Eq.~\eqref{AmentEq} to accurately calculate the RIXS phonon contribution.  We conduct this test using the O-K edge of acetone.  This choice is motivated by the availability of high quality experimental data \cite{Schreck2016} showing a clear and substantial phonon progression in the RIXS spectrum, the relative simplicity of the intermediate state that involves two degenerate anti-bonding electronic $\pi^*$ orbitals coupled to one vibrational mode (the C=O bond stretching), and the suitability of acetone for first-principles calculations (unlike strongly correlated materials).

To probe the interpretation of $g$ we evaluate the O-K RIXS spectrum of acetone with Eq.~\eqref{AmentEq} for three values of $g$ corresponding to electron-phonon coupling ($g_e$), core-hole--phonon coupling ($g_h$), and exciton-phonon coupling ($g_{eh}$).  To calculate these $g$ values we first use density functional theory to construct the ground-state and excited-state potential energy surfaces along the C=O bond stretching mode of gas-phase acetone.  The excited-state potential energy surface is calculated under the three respective conditions: ($g_e$) the addition of an electron to the lowest unoccupied molecular orbital (LUMO), ($g_h$) the introduction of a 1s core-hole on oxygen, and ($g_{eh}$) the addition of an electron to the LUMO and the introduction of an oxygen core-hole in order to simulate a core-valence exciton.  The four potential energy surfaces (including the ground-state (GS) surface) are presented in the inset of Fig.~\ref{modelRIXS}.  The excited-state forces for the three scenarios are obtained as the DFT Hellmann-Feynman forces using the ground-state equilibrium atomic positions, or equivalently by taking the derivative of the respective potential energy surfaces at the ground-state equilibrium bond length.  The coupling constants are related to the force constants by $M=\sqrt{\frac{\hbar}{2\mu\omega}}|F|$ or $g=\frac{F^2}{2\hbar\mu\omega^3}$ where $\mu$ is the reduced mass of the oxygen in acetone and $\omega$ corresponds to the respective excited-state vibrational frequencies (see Eq.~\eqref{XPSforce} for a general expression of the coupling constant).  The values of the forces are given in Table 1 of Appendix A along with numerical details of the calculations.

We next evaluate Eq.~\eqref{AmentEq} with each of the three coupling constants treated as fixed parameters.  The resulting phonon RIXS spectra are presented in Fig.~\ref{modelRIXS} compared to the experimental results of Schreck {\it et al.~}\cite{Schreck2016}.  The conclusion is immediately evident: the value of $g$ corresponding to the electron-phonon coupling case is unable to explain the experimental data, but using the value of $g$ corresponding to exciton-phonon coupling yields a result in close agreement with experiment.

The use of Eq.~\eqref{AmentEq}, which results from the canonical transformation of Eq.~\eqref{reducedHam}, is only strictly justified in the first two scenarios in which a single fermion is added (or removed) from the ground-state.  When we add an exciton, which can be viewed as a quasi-boson, each $c_i$ operator in Eq.~\eqref{reducedHam} is replaced by a product of two fermion operators and the canonical transformation no longer strictly holds \cite{Keldysh1968}.  However, in the low exciton density limit the canonical transformation is reasonably satisfied \cite{Agranovich1968, Grover1970}.

% Paragraph on computational schemes for e-ph RIXS
The agreement with experiment found when using $g_{eh}$ rather than $g_e$ can be rationalized by the observation that phonons in RIXS are created by the perturbation in the intermediate state, which consists of an exciton.  Our interpretation of RIXS as a probe of exciton-phonon coupling rather than electron-phonon coupling is implicitly supported by previous quantum chemistry or DFT based calculations of RIXS spectra of small molecules \cite{Saek2003, Sun2011, Couto2017, Ertan2017, VazdaCruz2017}.  These calculations evaluate the Kramers-Heisenberg equation for the RIXS cross section in the time-domain by time-evolving a nuclear wavepacket on a potential energy surface constructed by explicitly including the core-hole in the active subspace.  Results of such calculations agree with experiment to a high level of accuracy.  While not previously explicitly stressed, this indicates the important role played by the core-hole.  Additionally, it has long been recognized that the coupling constant probed in optical Raman spectroscopy relates to the intermediate state interaction of an optical exciton and a phonon \cite{Perebeinos2005, Kira2006, Hoffmann2002, Krauss1997, Antonius2017}.  Given the close analogy between the Raman and RIXS processes, it should not be entirely surprising that RIXS measures the coupling strength between an intermediate state core-level exciton and a phonon.  Lastly, we note that the concept of exciton-phonon coupling is not new.  In addition to the optical Raman work, exciton-phonon coupling has been described explicitly in calculations of the temperature dependence of optical absorption spectra \cite{Yarkony1976, Marini1994, Marini2008, Antonius2017}.

Only in the case that the core-hole is well screened and other excited-state effects may be neglected could the RIXS coupling constant approach the electron-phonon coupling value.  However, we expect that these conditions will generally not hold, even for the copper $L_{3}$ edge.  Tests for Cu$_2$O, presented in Appendix B, indicate that a Cu $2p$ core-hole is not particularly well screened and should contribute appreciably to the generation of phonons during a RIXS measurement.  Although RIXS measures exciton-phonon coupling rather than electron-phonon coupling we anticipate that it will be possible to disentangle the latter from the former, which will be the subject of future work.

\section{XPS : 1-particle spectral functions}
\label{XPSsect}

During the X-ray photoemission process, a core-level electron absorbs an X-ray photon and gets ejected from the material.  The resulting core-hole acts as a perturbation that can cause secondary excitations; in addition to the main quasiparticle peak, these satellites are referred to as the intrinsic contribution.  The photoelectron likewise acts as a brief perturbation when initially removed.  This gives the extrinsic effects.  The intrinsic and extrinsic effects may counteract, giving rise to an interference term.  For XPS, we ignore the role of the photo-electron and treat only the intrinsic contribution to vibrational excitations.

Within the above approximation, the XPS photocurrent is proportional to the core-hole spectral function, or the imaginary part of the core-hole ($\alpha$) Green's function.  Normally, one constructs the dressed Green's function $G_{\alpha}(\omega)$ from a bare Green's function $G_{\alpha}^{0}(\omega)$ and an approximate self-energy, $\Sigma_{\alpha}(\omega)$, through the Dyson equation $[G_{\alpha}(\omega)]^{-1}=[G_{\alpha}^{0}(\omega)]^{-1}-\Sigma_{\alpha}(\omega)$.  While this may be effective at renormalizing quasiparticle energies, for low-order approximations to the self-energy it typically yields poor spectral functions with weakened and misplaced satellite features \cite{Guzzo2011, Lischner2013}.  A better procedure, well suited to core-level spectroscopy \cite{Langreth1970, Nozieres1974, Hedin1980,  Gunnarsson1994,Aryasetiawan1996}, is to re-express the full Green's function as a cumulant expansion in the time domain $G_{\alpha}(t) = G_{\alpha}^{0}(t)e^{C_{\alpha}(t)}$ where $C_{\alpha}(t)$ is the cumulant.  The diagrammatic summation of this series, using the same approximation to the self-energy, implicitly recovers some vertex corrections neglected by the Dyson series and produces superior spectral functions.

The cumulant ansatz to the interacting Green's function originates in the linked-cluster expansion \cite{AbrikosovAAGorkovLP1975} and provides an exact solution with only a lowest order self-energy for the case of an isolated, deep core-hole \citep{Nozieres1969,Langreth1970}.  This is similar to the observation that the Holstein-type vibronic Hamiltonian can be solved exactly by a canonical transformation when electronic recoil is neglected.  The cumulant expansion is no longer exact when recoil becomes important, such as for valence level electrons and holes, nevertheless, its recent use in such cases has been promising \cite{Guzzo2011,Lischner2013,Kas2014, Zhou2015a,Nery2017}.  

First-principles work with the cumulant expansion has mainly been employed to describe plasmon satellites in conjunction with Hedin's GW self-energy (G is the single particle Green's function and W the screened Coulomb interaction).  This has proven to be quite successful at reproducing the valence-level photoemission spectrum of silicon \citep{Caruso2016,Borgatti2017,Caruso2015,Aryasetiawan1996,Guzzo2011,Onida2002} and has even reasonably reproduced the low energy plasmon satellites in SrVO$_{3}$ \cite{Gatti2013, Nakamura2016} and charge transfer effects in the XPS \cite{Kas2015} and XAS \cite{Kas2016} spectra of NiO, suggesting the applicability of the GW-cumulant approach may extend to moderately correlated materials.  Only very recently has the cumulant expansion been applied to the vibronic coupling problem \cite{Story2014,Gali2016,Nery2017}.  Here, one exchanges the plasmonic GW self-energy for the Fan-Migdal vibronic self-energy GD (D is the phonon propagator).  These studies focused on valence level photoemission; we are unaware of any application of vibronic cumulants to core-level spectroscopies. 

Working at a second-order approximation, the cumulant may be expressed in the time domain as
\begin{eqnarray}
&& C_{\alpha}(t,t^{\prime}) = \left [ G_{\alpha}^0(t,t^{\prime}) \right ]^{-1} \nonumber \\
&\times& \int_{t'}^t \int_{t'}^t  dt_1 dt_2 \, G_{\alpha}^0(t,t_1) \Sigma_{\alpha} (t_1,t_2) G_{\alpha}^0(t_2,t^{\prime}) \, .
\label{core-cumulant}
\end{eqnarray}
We approximate the phonon contribution to the core-hole self-energy with the $2^{nd}$ order Fan-Migdal term 
~
\begin{equation}
\label{sigma-xps}
\Sigma_{\alpha}(t^{\prime},t)= i\sum_{{\bf q}\lambda} |M_{\alpha}^{{\bf q}\lambda}|^2 G^0_{\alpha}(t^{\prime},t) D_{{\bf q}\lambda}(t^{\prime},t) \, ,
\end{equation}
which neglects the Debye-Waller contribution.  The localized nature of the core-hole in XPS permits the neglect of recoil, which significantly simplifies the problem as was shown by Nozi\`{e}res and De Dominicis \cite{Nozieres1969} and Langreth \cite{Langreth1970} for plasmons, and by Dunn \cite{Dunn1975} and Gunnarsson {\it et al.~}\cite{Gunnarsson1994} for phonons.  The core-hole Green's functions may be factored out of the integrals in Eq.~\eqref{core-cumulant} to give
~
\begin{equation}
\label{norecoilcumulant}
C_\alpha(t,t^{\prime})=-\sum_{{\bf q}\lambda} |M^{{\bf q}\lambda}_{\alpha}|^2\int^t_{t^{\prime}}\int_{t^{\prime}}^t d\tau d\tau' D_{{\bf q}\lambda}(\tau,\tau') \theta(\tau-\tau^{\prime}) \, .
\end{equation}
The two terms required to evaluate the core-hole cumulant are the phonon Green's function in the presence of the core-hole, $D_{{\bf q}\lambda}$, and the core-hole--phonon coupling parameter, $M_{\alpha}^{{\bf q}\lambda}$.  These quantities may be obtained with density functional theory or density functional perturbation theory.  For small molecules, such as we consider below, {\it ab initio} molecular dynamics (AIMD) is also a convenient way to obtain both quantities.
 
Within the AIMD approach, the time-ordered phonon Green's function can be expressed as the autocorrelation function $iD_{{\bf q}\lambda}(t) =  \bra{0}TB_{{\bf q}\lambda}(t)B_{{\bf q}\lambda}(0)\ket{0}$ where $B_{{\bf q}\lambda}(t)=b_{-{\bf q}\lambda}^{+}(t) + b_{{\bf q}\lambda}(t)$ is the displacement operator for a phonon of mode $\lambda$ and momentum ${\bf q}$.  The classical equivalent, suitable when the nuclei are treated classically within MD, is
~
\begin{equation}
D_{{\bf q}\lambda}(t,t^{\prime}) = -i\sum_{ij} Q_{i}^{{\bf q}\lambda}(t) Q_{j}^{{\bf q}\lambda}(t^{\prime}) e^{-i{\bf q} \cdot ({\bf R}_i-{\bf R}_j)} 
\end{equation}
in which $Q_{i}^{{\bf q}\lambda}(t) = \sqrt{\frac{\mu_i \omega_{{\bf q}\lambda}}{2N\hbar}} \delta \vec{R}_{i} \cdot \vec{\xi}_{i}^{{\bf q}\lambda}$ is the normalized atomic displacement projected along the phonon normal mode vector $\xi_{i}^{{\bf q}\lambda}$ for atom $i$.   $N$ is the number of atoms in the unit cell and $\mu$ is the average reduced mass of the unit cell.  The electron-phonon coupling is given by
~
\begin{equation}
\label{XPSforce}
M_{\alpha}^{{\bf q}\lambda} = \sum_{j} e^{-i{\bf q} \cdot {\bf R}_j} \sqrt{\frac{\hbar}{2N\mu_j \omega_{{\bf q}\lambda}}} \vec{F}_{j}({\bf R}_{j},t=0) \cdot \vec{\xi}^{{\bf q}\lambda}
\end{equation}
with $F_j$ the force on atom $j$ at position ${\bf R}_j$ at $t=0$.

\subparagraph{Numerical results:}

\begin{figure}[h]
\begin{minipage}[h]{1\linewidth}
\center{\includegraphics[width=1\linewidth]{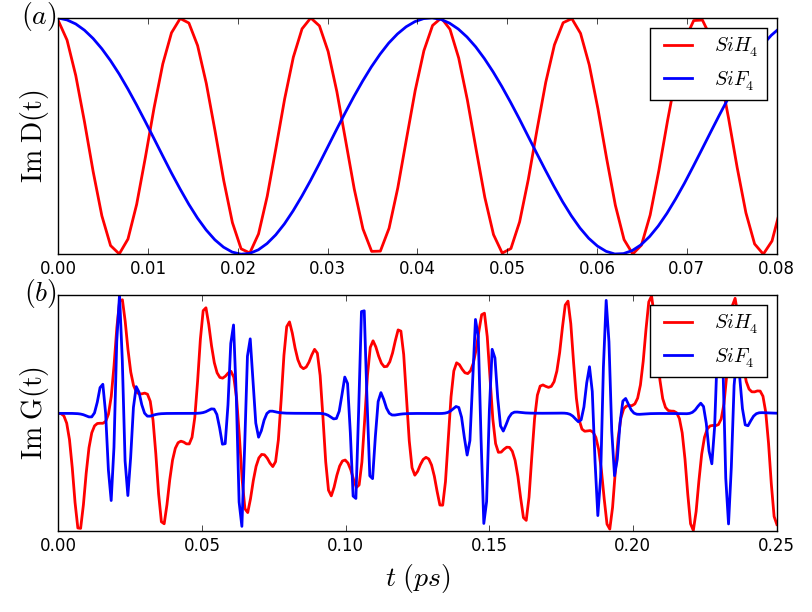}}	
\end{minipage}
\hfill
\begin{minipage}[h]{1\linewidth}
\center{\includegraphics[width=1\linewidth]{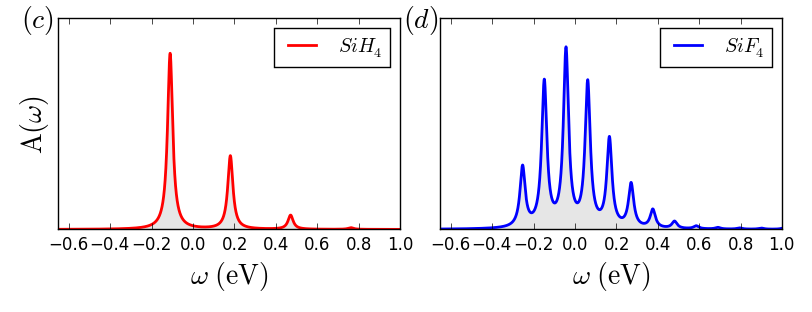}}
\end{minipage}

\caption{Construction of the core-hole vibronic spectral function. $(a)$ The imaginary part of the ligand displacement-displacement correlation function for SiH$_{4}$ (red) and SiF$_{4}$ (blue) obtained from {\it ab inito} molecular dynamics simulations in the presence of a Si 2$p$ core-hole.  $(b)$ The imaginary part of the full core-hole Green's function for SiH$_{4}$ (red) and SiF$_{4}$ (blue) in the time domain.  Silicon 2$p$ core-hole spectral function due to vibrational interactions for $(c)$ SiH$_{4}$ (red) and $(d)$ SiF$_{4}$ (blue).  The first peak of each spectral function is displaced from zero by $-M^2/\omega$ as discussed in the text.}
\label{fig-core-response}
\end{figure}

We consider the Si 2$p$ core-hole vibrational coupling in the X-ray photoemission spectra of SiH$_4$ and SiF$_4$ molecules.  The main motivations for this choice are the availability of high quality experimental data \cite{Thomas2002} showing clearly distinguishable phonon sidebands and the suitability of the system to first-principles methods.  To construct the electron-phonon self-energy from Eq.~\eqref{sigma-xps} we generate the phonon Green's function and the core-hole phonon coupling constant from a molecule dynamics simulation in which we switch on a core-hole in the silicon 2$p$ level at $t=0$.  Tracking the positions of ligand atoms (H or F), we observe that only the $A_1$ breathing mode is excited by the core-hole, which is expected from symmetry considerations.  We therefore express the dynamics in terms of this breathing mode.  Figure \ref{fig-core-response}a shows the imaginary part of the breathing mode phonon Green's function in the time domain constructed as the autocorrelation function of the normal mode coordinate.

An advantage of using the AIMD autocorrelation function to construct the vibrational Green's function is that it naturally captures any anharmonic contribution present, though the response in this case is nearly harmonic.  Since the observed response is harmonic and $t=0$ corresponds to a maximum in the displacement amplitude, we can use the AIMD forces at $t=0$ to obtain the core-hole phonon coupling constant according to Eq.~\eqref{XPSforce}.  The core-hole self-energy is evaluated by Eq.~\eqref{sigma-xps}, which then gives the cumulant by Eq.~\eqref{core-cumulant} and finally the dressed core-hole Green's function.  The imaginary part of the core-hole Green's function in the time domain is shown in Fig.~\ref{fig-core-response}b.  Fourier transforming to frequency space gives the core-hole spectral functions, shown in Figs.~\ref{fig-core-response}c-d for SiH$_4$ and SiF$_4$, respectively.  The Lorentzian width of the features in Figs.~\ref{fig-core-response} originates from the addition of a finite CVV Auger core-hole lifetime \cite{Larkins1994, Thomas2002} to the bare core-hole Green's function.

A few observations about the spectral functions are worth mentioning.  Most immediately, the spectral functions show clear vibrational sidebands at multiples of the phonon frequency.  The degree of asymmetry of each series depends on the coupling strength.  For SiH$_4$, the coupling constant is relatively weak ($g=0.25$) and only three features are seen in the spectral function with monotonically decreasing intensities, giving a very asymmetric structure.  The situation is quite different for SiF$_4$, which exhibits intermediate coupling ($g=2.4$); numerous vibronic features are observed with non-monotonically varying intensities, giving a more symmetric overall form.

Due to the vibronic coupling, there is a well known shift \citep{Langreth1970,Nery2017} of the first vibrational peak to lower energy by $-M_\lambda^2/\omega_\lambda$ with respect to the bare-particle peak position.  For this reason, the leading, zero phonon peak in each spectral function appears at negative energy.  For weak coupling (SiH$_4$) this shifts the overall spectral weight to slightly lower energy, while in the intermediate coupling case (SiF$_4$), the overall spectral weight is barely displaced.  Often, vibrational features are not experimentally resolved and contribute to a spectral width that is generically interpreted.  Neglect of this vibronic shift can lead to misquantification of bare-particle energies, which can confuse comparison between experiment and calculation.

In more general cases, the above observations may be precisely quantified with the aid of the boson excitation spectrum 
~
\begin{equation}
\label{beta-general}
\beta_{\alpha}(\omega) = \frac{1}{\pi} \left | {\rm Im}  \, \Sigma_{\alpha}(\omega+\epsilon_\alpha) \right | \, .
\end{equation}
Physically, the response of a system to the perturbation caused by a core-excitation can be described in terms of secondary bosonic excitations (phonons in this case).  The intensity with which bosons of energy $\omega$ are generated by the perturbation is given by $\beta_{\alpha}(\omega)$.  The energy shift and renormalization factor of the quasiparticle peak can be expressed in terms of the excitation spectrum, respectively, as
~
\begin{equation}
\Delta E_{\alpha} = \int d\omega \frac{\beta_{\alpha}(\omega)}{\omega} \;\;\;\; {\rm and } \;\;\;\; Z_{\alpha}= \exp \left[ \int d\omega \frac{\beta_{\alpha}(\omega)}{\omega^2} \right] \, .
\end{equation}
The boson excitation spectrum gives an alternate expression for the cumulant (at the 2$^{nd}$ order)
\begin{equation}
\label{cumulant_a}
C_{\alpha}(t) = \int d\omega \frac{\beta_{\alpha}(\omega)}{\omega^2} \left [ e^{-i\omega t} +i\omega t - 1\right ] 
\end{equation}
that can be convenient for frequency-domain calculations.

\begin{figure}
\begin{minipage}[h]{1.0\linewidth}
\center{\includegraphics[width=1\linewidth]{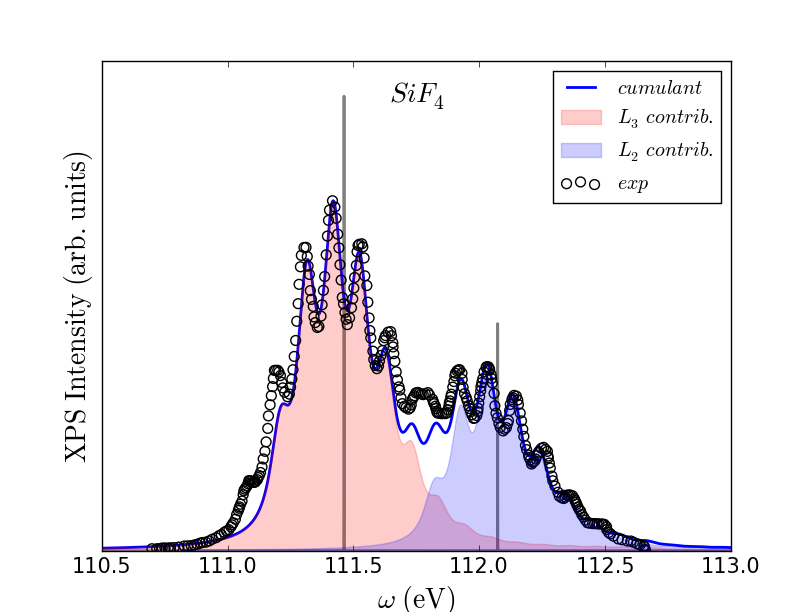}}    
\end{minipage}
\hfill
\begin{minipage}[h]{1.0\linewidth}
\center{\includegraphics[width=1\linewidth]{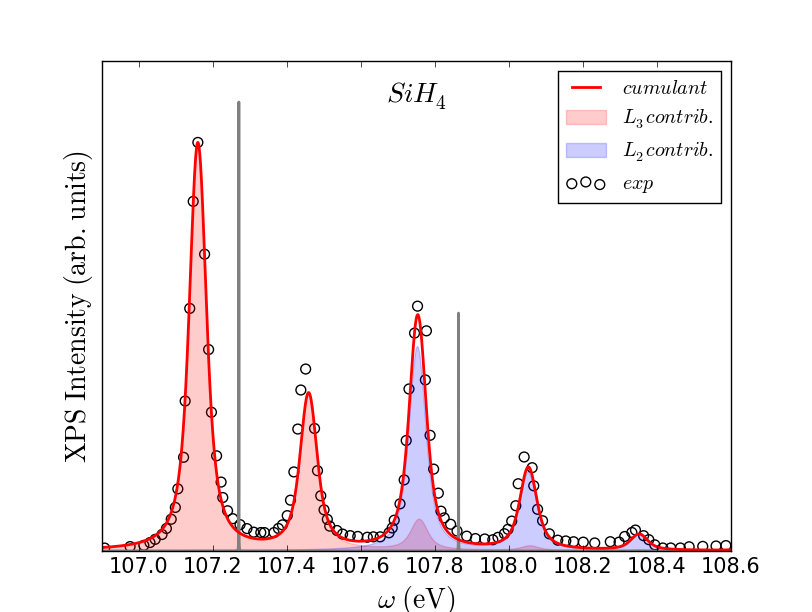}}
\end{minipage}

\caption{Calculated XPS spectra of SiH$_4$ (top) and SiF$_4$ (bottom) compared to experimental results \citep{Thomas2002} (open symbols). Shaded curves represent the individual $2p_{3/2}$ (red) and $2p_{1/2}$ (blue) contributions. Vertical gray lines indicate the positions of electronic peaks in the absence of electron-phonon interactions.}
\label{xpsfig}
\end{figure}

Turning finally to the XPS spectra, the silicon $2p$ orbitals are split by spin-orbit coupling into $2p_{3/2}$ and $2p_{1/2}$ levels, which are separated by about 0.6 eV \cite{Sankari2003}.  To compare our calculations with experimental results \cite{Thomas2002}, we obtain the full silicon $2p$ XPS signal by convolving each spectral function with a bare core-hole spectrum that consists of two delta functions (for the $2p_{3/2}$ and $2p_{1/2}$ levels) that have a 2:1 intensity ratio.  This comparison is presented in Fig.~\ref{xpsfig}.  An additional linewidth, beyond the Lorentzian core-hole broadening, was added by further convolving with a Gaussian to account for the experimental resolution \cite{Thomas2002}.  We note that the experimental observation of phonon sidebands in XPS further demonstrates the ability of a core-hole to create phonons.

\section{XAS : 2-particle spectral functions}
\label{xas}

It is often a good approximation to describe the X-ray absorption event with a two-particle picture in which a photon creates a core-hole and a photo-electron (or annihilates the pair for the X-ray emission case).  Evaluation of the two-particle Green's function presents a significant challenge beyond the one-particle Green's function as one must now treat the interaction between the photo-electron and core-hole in addition to their interactions with the lattice.  Following the recent suggestion of Antonius and Louie \cite{Antonius2017}, we first solve the electronic problem for the two-particle excitonic states, and then dress the excitons with phonons.  This separation of the electronic interaction kernel from the electron-lattice problem is justified by the significantly different time scales of the interactions.  Whereas Antonius and Louie considered vibronic effects on optical excitons at the Dyson-Migdal level, we treat the phonon contribution to the core-level exciton within the Migdal-cumulant approximation.

We generate the purely electronic two-particle  Green's function $L(\omega)$ and corresponding excitation spectrum $\mu^{0}(\omega)$ by solving the Bethe-Salpeter equation (BSE) \cite{Salpeter1951}.  The BSE describes the electron and hole quasi-particles at the density functional theory level, typically with GW self-energy corrections.
Near threshold, the electron and hole can pair into an exciton state by interacting with each other through the screened direct and bare exchange interactions, which are treated by summing ladder diagrams.  Each excitonic state $\xi$ will be a mixture of electron-hole pairs $\ket{\phi_{\xi}} = \sum_{n{\bf k}\alpha} A_{n{\bf k}\alpha}^{\xi} \ket{n{\bf k};\alpha}$ and thus the exciton creation operator is defined in terms of electron and hole operators as $a^+_{\xi} = \sum_{n{\bf k}\alpha} A_{n{\bf k}\alpha}^{\xi} c^+_{n{\bf k}}c_{\alpha}$.  Details of the BSE as it applies to optical and core-level excitations can be found in many references \cite{ Rohlfing1997, Rohlfing1998,Shirley1998, Albrecht1998} and reviews \cite{Onida2002}.

Using this electronic solution as a starting point, we now treat the interaction between the purely electronic excitons and phonons. Denoting the phonon-dressed exciton Green's function as $\Lambda(\omega)$, the X-ray absorption coefficient is
~
\begin{equation}
\label{mu}
\mu(\omega) = -\frac{1}{\pi}\sum_{\xi,\xi'} (d_\xi)^* d_{\xi'} {\rm Im} \bra{0}a_{\xi'} \Lambda(\omega+\epsilon_i)a_{\xi}^+ \ket{0} \,
\end{equation}
where the matrix elements of the photon operator $\Delta$ are given by $d_\xi=\bra{\psi_\xi}\Delta\ket{0}$.  The ground state wave-function $\ket{0}$ contains no excitons or phonons in the limit of zero temperature. The vibronic wavefunction $\ket{\psi_\xi}=a_{\xi}^+ \ket{0}=\prod_\nu\ket{\phi_\xi}\ket{n_\nu=0}$ contains both vibrational $\ket{n_{\nu}}$ and electronic $\ket{\phi_{\xi}}$ parts, but is a solution of $H_0$ (without an electron-phonon interaction).  (Here we have introduced the short-hand notation $\nu = \{ {\bf q}, \lambda \}$ for the phonon mode and momentum.)  The photon operator does not directly create phonons within the ground-state vibrational basis.  Rather, phonons are generated by the electron-lattice interaction during the propagation of the exciton.

The main challenge is to evaluate the dressed exciton propagator $\Lambda(\xi',\xi,\omega+\epsilon_i)=\bra{0}a_{\xi'} \Lambda(\omega+\epsilon_i)a_{\xi}^+ \ket{0}$.  In the absence of the electron-phonon interaction (or more complex electronic interactions not captured by the BSE) $\Lambda (\xi', \xi,\omega)$ will be diagonal ($\xi=\xi'$) and will reduce to the bare exciton propagator $L(\xi, \omega)$.  Exciton-phonon scattering can mix the pure excitonic states through intraband scattering, which we include in the propagation of the exciton.  In principle, scattering can also give rise to non-diagonal elements of the propagator $(\xi \neq \xi')$, however, the non-diagonal elements typically involve interband scattering, which will be negligible except near band crossings.  Therefore, we restrict our effort to just the diagonal contributions.  

Within the cumulant notation, the phonon-dressed exciton Green's function may be expressed in terms of the bare exciton Green's function as
~
\begin{equation}
  \label{2cumulant}
\Lambda(\xi,t) = L(\xi,t) e^{C(\xi,t)} \ .
\end{equation}
Here we treat the exciton as an effective quasi-particle that interacts with phonons.  If the exciton is restricted to an isolated level and recoil during phonon scattering may be neglected, evaluation of the cumulant in Eq.~\eqref{2cumulant} follows similarly to the procedure presented in the XPS section.  However, this will not be the case in general.

Analogously to Eq.~\eqref{core-cumulant}, a second-order approximation to the exciton cumulant may be written as
\begin{align}
\label{xasrecoil}
&C(\xi,t)=[L(\xi,t)]^{-1} \nonumber\\
&\times \int_0^t \int_0^t d\tau d\tau' L(\xi,t-\tau) \Sigma(\xi,\tau-\tau') L(\xi,\tau)  \ .
\end{align}
The exciton self-energy $\Sigma_\xi$ contains a generalized Fan-Migdal like term 
\begin{equation}
\label{sigma-xas}
\Sigma^{FM}(\xi,t) = i\sum_{{\nu,\xi_1}} [M_{\xi\xi_1}^{\nu}]^2  L(\xi_1,t)D_{\nu}(t) \, 
\end{equation}
for the interaction of a bare exciton with a phonon.  Here, there is an important difference with respect to the XPS core-hole self-energy in Eq.~\eqref{sigma-xps}.  Equation \eqref{sigma-xas} contains a summation over exciton states, which accounts for intraband scattering during the propagation.  Consequently, it is not possible to factor out the propagator lines in Eq.~\eqref{xasrecoil} as was done for the XPS problem to arrive at Eq.~\eqref{norecoilcumulant}, which was effectively an exact solution.

\begin{figure}[htb]\centering
\center{\includegraphics[width=1\linewidth]{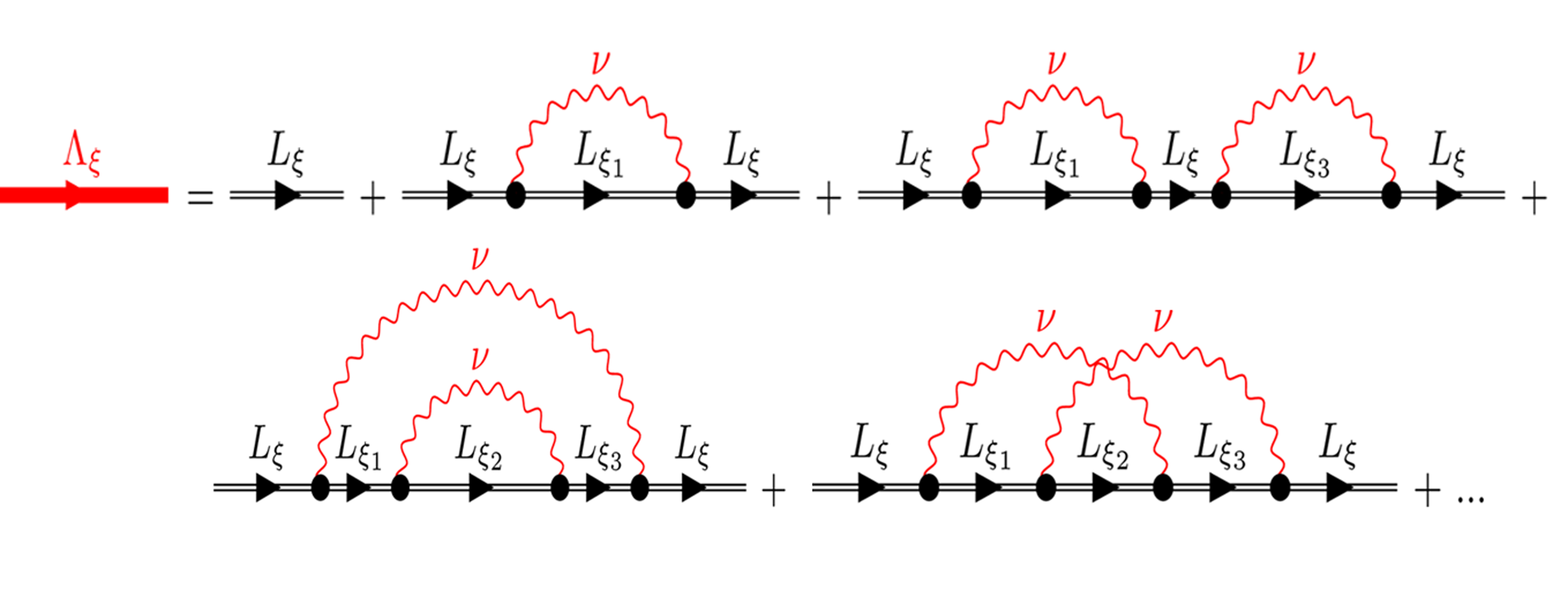}}	
\caption{Diagrammatic series included in the cumulant expansion for the fully interacting exciton Green's function $\Lambda_\xi$ (in red).  The 2-particle bare exciton propagator $L_\xi$ is represented by the double black lines while the oscillating red curves indicate phonons.  The cumulant expansion accounts for self-consistency terms, such as rainbow diagrams ($4^{th}$ diagram), and approximately includes vertex corrections ($5^{th}$ diagram) \cite{Hedin1980}.  Higher order replicas are also included.}
\label{diagramXAS}
\end{figure}

Even though the cumulant expansion is no longer exact when phonons can scatter the exciton between different states, a reasonable result can still be obtained from the second order approximation through exponential re-summation \cite{Mahan1990, Kas2014}.  Expanding the cumulant in powers of the exciton-phonon interaction as $C(t) = \sum_m C_m(t)$, we see that even stopping at second-order in the cumulant, the exponential form of the exciton propagator $\Lambda(t) = L(t) \exp[C_2(t)]$ will accumulate terms to infinite order.  Beyond generating multiple replicas of the Fan-Migdal diagram, the second-order cumulant will also approximately reproduce diagrams that include vertex corrections such as the last one in Fig.~\ref{diagramXAS}. This partial re-summation is a common approximation for the cumulant  \cite{Kas2014, Story2014, Kas2015, Nery2017} for both phonon and electronic contributions.

\subparagraph{Exciton-phonon coupling:}

The electron-lattice matrix elements $M_{\xi\xi^{\prime}}^{\nu}$ are evaluated as an exciton-phonon scattering process rather than separate electron-phonon and core-hole--phonon scattering events.  We construct the exciton-phonon coupling constants by generalizing the frozen-phonon procedure presented by Tinte and Shirley \cite{Tinte2008} and Gilmore and Shirley \cite{Gilmore2010} to study the vibrational contribution to the XAS linewidth of SrTiO$_{3}$.  This defines the force constants
\begin{equation}
\label{BSE-forces}
F_{\xi\xi^{\prime}}^{\nu} =\left . -\partial_{Q_{\nu}} \left [ E_{tot}^{GS}(Q_{\nu}) + \bra{\xi^{\prime}} H_{BSE}(Q_{\nu}) \ket{\xi} \right ] \, \right |_{\delta Q_{\nu}=0}
\end{equation}
as derivatives of the excited-state total energy with respect to given atomic displacements. This expression partitions the atomic position dependent excited-state total energy into the sum of the ground-state potential energy surface $E_{tot}^{GS}(Q)$ and the energy separation between the ground- and excited-state PES ($H_{BSE}(Q)$).  In the above expression $\xi$ and $\xi^{\prime}$ are excitonic eigenstates of the equilibrium lattice and Eq.~\eqref{BSE-forces} accounts also for the scattering between exciton states by phonons. The exciton-phonon coupling  constants are obtained from the force constants as $M^\nu_{\xi\xi'}=\sqrt{\frac{\hbar}{2\mu \omega_\nu}}F^\nu_{\xi\xi'}$.  We anticipate that it should be possible in the general case of a periodic solid to construct the exciton-phonon coupling constants without the need for supercells in a manner analogous to density functional perturbation theory.

To evaluate the Fan-Migdal exciton self-energy we need the phonon Green's function in addition to the exciton-phonon coupling constants.  Rather than calculate explicitly the full response of the lattice to the creation of an exciton we assume that the lattice response is harmonic with a frequency extracted from the excited-state potential energy surface, which is obtained during the calculation of the force constants.

\begin{figure}
\center{\includegraphics[width=1\linewidth]{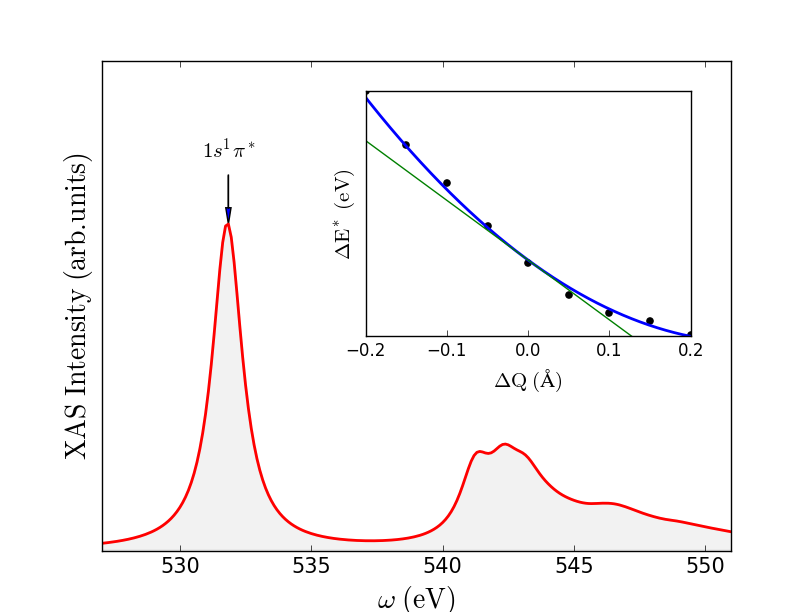}}	
\caption{Calculation of the O-K XAS of acetone.  The arrow indicate the $1s^{1}\pi^{*}$ resonance.  The inset shows the $1s^{1}\pi^{*}$ excited-state PES with respect to the C-O bond length constructed as the sum of the ground-state PES and the BSE excitation energy.  The tangent line at the ground-state C-O equilibrium bond length gives the excited-state force while the excited-state vibrational frequency is obtained from the quadratic fit (blue curve).}
\label{forcecalc}
\end{figure}

\begin{figure*}
\begin{minipage}[h]{1\textwidth}
\center{\includegraphics[width=1\textwidth]{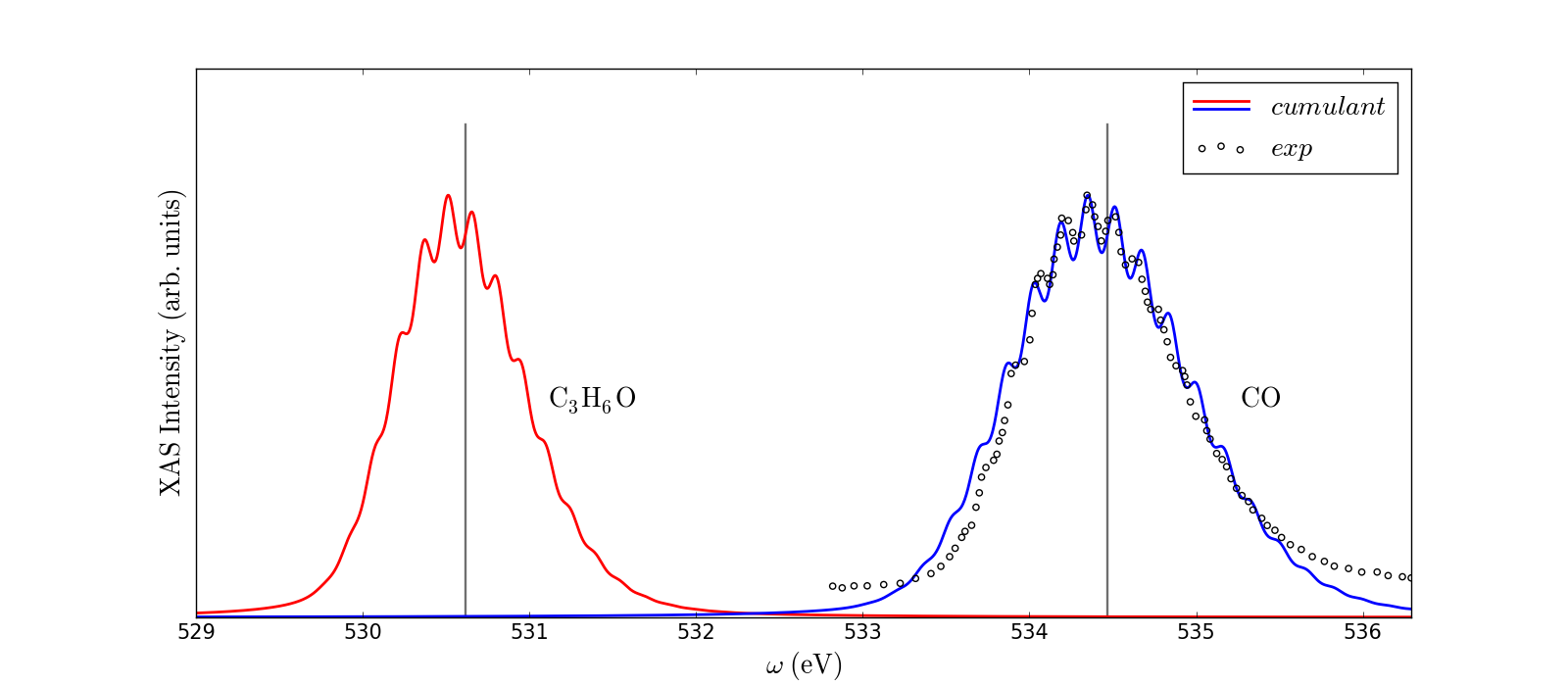}}
\end{minipage}
\caption{Vibrational contribution to the X-ray absorption spectra of the O K-edge of acetone (C$_{3}$H$_{6}$O) and CO.  Theoretical spectra (colored curves) are compared to experiment (open symbols).  Gray vertical lines indicate the energy positions of the purely electronic quasiparticle spectra.  The energy shift and degree of symmetry of the vibronic spectra depend on the exciton-phonon coupling strength.  Experimental results are taken from  P\"uttner ${et \ al.}$ \cite{co}.}
\label{fig-xas}
\end{figure*}

\subparagraph{Numerical results:}

To demonstrate the above methodology we calculate the vibrational contributions to the N K-edge of the N$_{2}$ molecule and to the O K-edge of gas-phase acetone and CO.  We select these three molecules because for each the problem can be well approximated as a single exciton state interacting with a single vibrational stretching mode, and because high resolution experimental data exists showing multiple phonon sidebands. The paragraphs below refer specifically to acetone unless otherwise stated.

The acetone O-K XAS consists of an isolated feature at 531.5 eV that corresponds to the excitation of an oxygen 1$s$ electron into a $\pi^{*}$ anti-bonding orbital between the oxygen and nearest carbon atom, and a broad continuum at higher energy.  We obtain the purely electronic absorption spectrum by solving the Bethe-Salpeter equation with all atoms fixed at their equilibrium positions using the OCEAN code \cite{Vinson2011,Gilmore2015}.  Figure \ref{forcecalc} compares this calculation with the experimental spectrum.  We focus on the $1s^{1}\pi^{*}$ feature at 531.5 eV.  To evaluate the exciton-vibron force constant we perform the numerical derivative in Eq.~\eqref{BSE-forces} explicitly by repeating the BSE calculation several times while moving the oxygen atom in order to make incremental adjustments to the C-O bond length.  The excited-state potential energy surface, constructed as the sum of the ground-state PES and the BSE excitation energy, is given in the inset of Fig.~\ref{forcecalc}.  The derivative of the excited-state potential energy surface at the ground-state equilibrium bond length gives an exciton-vibrational force constant of $F=-7.6$ eV/${\rm \AA}$ for the bond stretching mode.  This is equivalent to a value of $M=0.35$ eV or $g=5.4$.  This compares very well to the value of $F=-7.7$ eV/${\rm \AA}$ obtained from the constrained DFT calculation presented in Sec.~\ref{sec:intro}. For CO molecule this leads to the slightly higher force of $F=-13.7$ eV/${\rm \AA}$ and stronger coupling  $M=0.6$ eV and $g=13$. The constrained DFT calculations for CO molecules give close value of the excited state force $F=-11.7$ eV/${\rm \AA}$. The same procedure was done for the N$_2$ molecule resulting in the force constants of  $F= -6.7$ eV/${\rm \AA}$, corresponding to $M=0.24$ eV and $g=0.92$, again agreeing well with the constrained DFT result of $F= -7.0$ eV/${\rm \AA}$. 

The phonon Green's function was modeled in the harmonic limit using the frequency found from a quadratic fit to the excited-state potential energy surface.  For acetone this gave $\omega=0.15$ eV,  $\omega=0.16$ eV for CO and for N$_2$ $\omega=0.25$ eV, though in the present case the frequencies can be clearly observed experimentally as the energy separation between phonon sidebands.  Anharmonic vibrational responses could be obtained through atomic displacement autocorrelation functions, as done for XPS, generated by excited-state molecule dynamic simulations.  However, beyond the challenge of such calculations, in many periodic systems the harmonic response will be sufficient. 

\begin{figure}
\begin{minipage}[h]{1\linewidth}
\center{\includegraphics[width=1\linewidth]{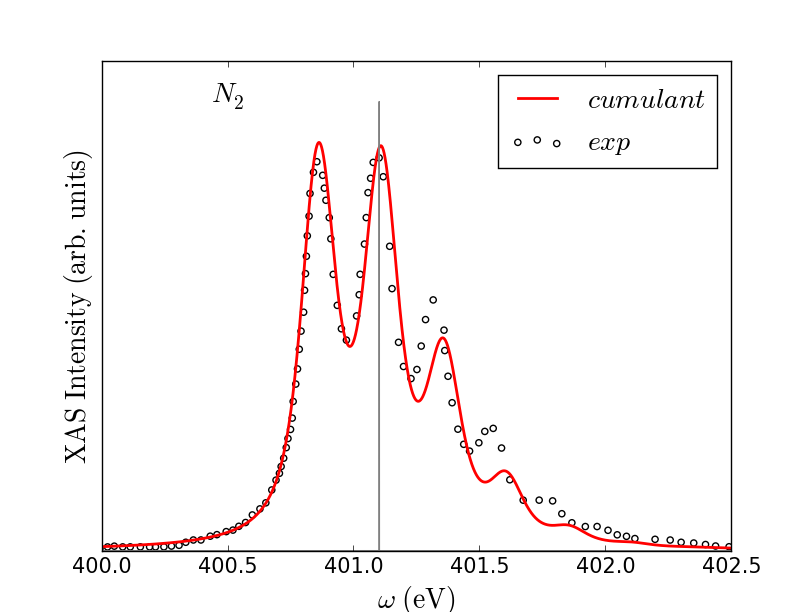}}	
\end{minipage}
\caption{Vibrational contribution to the X-ray absorption spectra of the K-edge of N$_{2}$.   Gray vertical line indicate the energy positions of the purely electronic quasiparticle spectra.    Experimental results are taken from Chen $et \ al.$ \cite{Chen1989}.}
\label{fig-xasn2}
\end{figure}

With the exciton-phonon coupling constant and the phonon Green's function, we evaluate the exciton self-energy ${\Sigma^{FM}_\xi}$ in Eq.~\eqref{sigma-xas}.  The exciton cumulant is then formed by Eq.~\eqref{xasrecoil} and the imaginary part of the resulting full exciton Green's function gives the effective XAS spectral function.  Since the two cases we are considering -- the oxygen $1s^{1}\pi^{*}$ resonance of acetone and the N-K edge of N$_2$ -- involve excitations into isolated levels, the exciton spectral function is effectively equivalent to the XAS signal (the pure electronic spectrum can be approximated as a single Lorentzian with core-hole lifetime broadening).  Additional instrumental broadening was added using a HWHM corresponding to the original experimental work \cite{Chen1989,co}.  The final spectra are compared to the experimental data in Figs.~\ref{fig-xas},\ref{fig-xasn2}.

As with the XPS results, the energy shift of the quasiparticle peak and the degree of symmetry of the overall spectral shape depend on the exciton-phonon coupling strength.  N$_2$ has relatively weak coupling ($g \sim 1$) and its spectrum shows noticeable asymmetry while the coupling strengths of acetone and CO are larger ($g \sim 5- 10$) and the spectrum is more symmetric.  The overall agreement between the calculation and experiment  is quite good while deviations in the intensities of higher order satellites are noticeable for N$_2$ and CO.  This could be an indication of anharmonicity of the excited-state vibrations that we have neglected.

\section{phonon contribution in RIXS}
\label{RIXS}
The cross-section for resonant inelastic X-ray scattering  was introduced in Sec.~\ref{sec:intro}.  In the present work, we restrict our attention to the so called quasi-elastic line.   In this limit, the system returns to the electronic ground state at the end of the scattering process, leaving only vibrational excitations such that the final state propagator, $D(\omega_{loss})$, is the ground-state phonon Green's function.  However, the results obtained in this section could be generalized to final states with a combination of electronic and vibrational excitations.

The RIXS cross-section can be written in terms of the interacting two particle propagator $\Lambda(\omega)$ without explicit summation over all intermediate states as
\begin{eqnarray}
\label{rixscross1}
\sigma(\omega_i,\omega_{loss})&=&\sum_f \left | \bra{\psi_f}\Delta_{o}^+\Lambda(\omega_i)\Delta_{i}\ket{\psi_i} \right |^2 \\ \nonumber
&\times& \delta(\epsilon_{fi}-\omega_{loss}) \, \, .
\end{eqnarray}
The photon operator $\Delta_i$ ($\Delta_o$) corresponds to the incoming (outgoing) photon.  Within the delta function, $\epsilon_{fi} = \epsilon_f - \epsilon_i$ is the energy difference between the initial and final states.  The final state vibronic wavefunction ($\ket{\psi_f}=\ket{\phi_i}\ket{n^{(f)}_{\nu}}$) differs from the initial wavefunction ($\ket{\psi_i}=\ket{\phi_i}\ket{n^{(i)}_{\nu}}$) only by the number of phonons in each state.  Therefore, we replace the summation index $f$ by the final-state vibrational levels, $n_{\nu}^{(f)}$, of each mode $\nu=\{ {\bf q}, \lambda \}$.  There is no summation over initial states since we again assume the low temperature limit that the initial state is the electronic ground-state with all phonons in the zero oscillator level.  For this reason, we will not explicitly include the $f$ superscript hereafter.

If we momentarily limit our consideration to a single phonon mode $\nu^{\prime}$ for notational simplicity, we can write the RIXS amplitudes in the excitonic basis as 
\begin{widetext}
\begin{equation}
\label{matrixelement}
\bra{n_{\nu^{\prime}}}\bra{\phi_i}\Delta_{o}^*\Lambda(\omega_i)\Delta_{i}\ket{\phi_i}\ket{0}= 
\sum_{\xi_1,\xi_2}(d^o_{\xi_2})^* d^i_{\xi_1} \Lambda^{(n)}(\xi_1,\xi_2, \nu^{\prime}, \omega_i) \, .
\end{equation}
For later convenience, on the right-hand side we separate the oscillator level $n$ (in the superscript of $\Lambda$) from the mode index $\nu^{\prime}$ (within the parentheses of $\Lambda$).  We can then express the RIXS cross section as a summation over $n$ phonon contributions $\Lambda^{(n)}$ as
\begin{equation}
\sigma(\omega_i,\omega_{loss}) = -\frac{1}{\pi} {\rm Im} \, \sum_{n}  | \sum_{\xi_1,\xi_2} (d^o_{\xi_2})^* d^i_{\xi_1} \Lambda^{(n)}(\xi_1,\xi_2,\nu', \omega_i)|^{2} D^{n}_{\nu'}(\omega_{loss}) \ .
\label{rixsvib}
\end{equation}
\end{widetext}
The final-state summation will be limited to a manageable range because the phonon contribution to RIXS at the quasielastic line will contain only a small number of observable phonon peaks.

\begin{figure}[htb]\centering
\center{\includegraphics[width=1\linewidth]{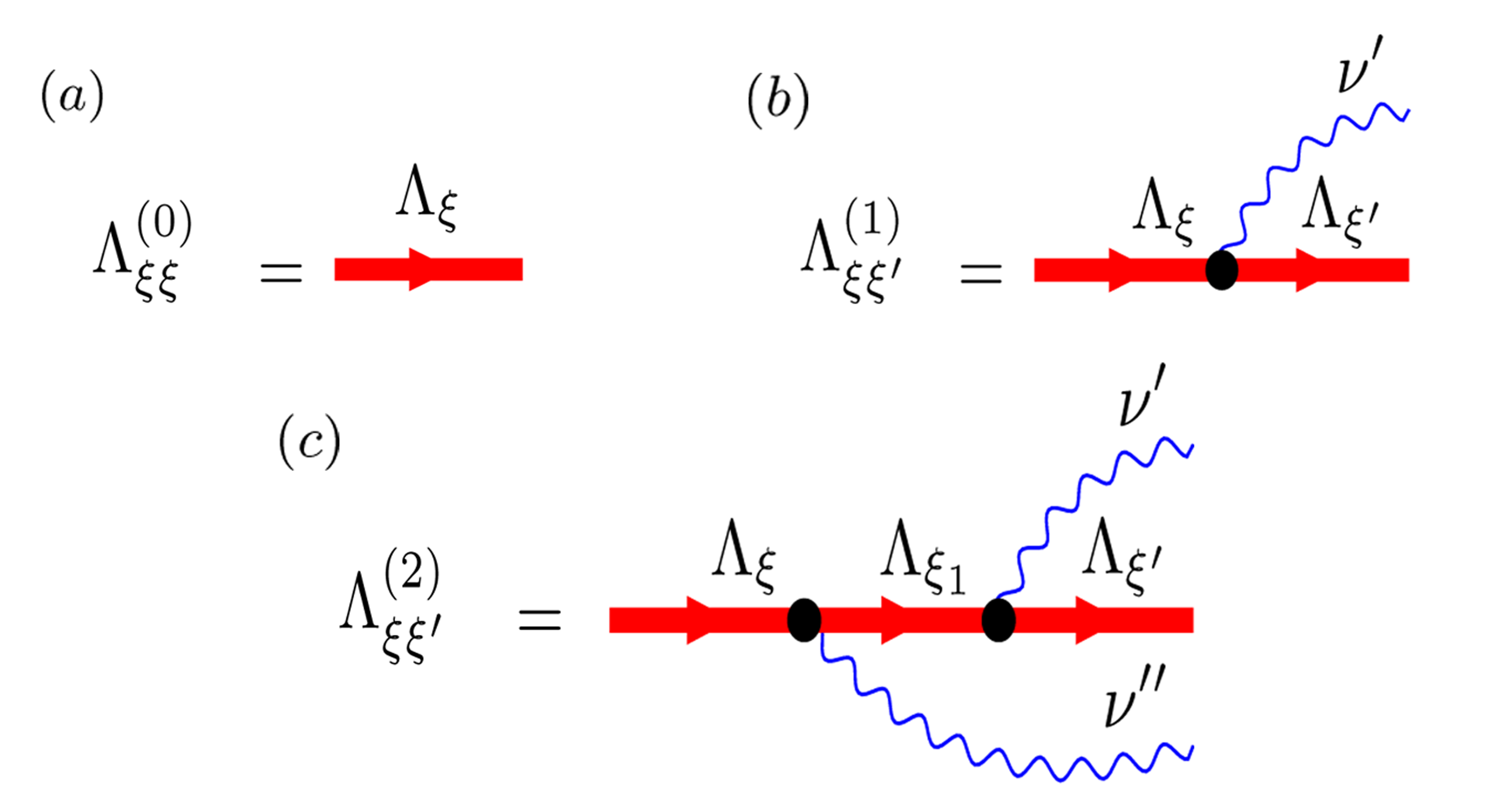}}	
\caption{Diagrams for the RIXS amplitudes for $(a)$ zero, $(b)$ one, and $(c)$ two real (final-state) phonons.  Heavy red lines indicate phonon-dressed exciton propagators and blue oscillating curves represent final-state phonons.}
\label{norecoil_d1}
\end{figure}

The RIXS amplitudes $\Lambda^{(n)}$ can be represented by the diagrams in Fig.~\ref{norecoil_d1}.  The basic element of the diagrams is $\Lambda_{\xi}$, which gives the full contribution in the $n=0$ case where the final state contains no phonons.  This is the same as the XAS phonon-dressed exciton propagator given by Eq.~\eqref{2cumulant} in Sec.~\ref{xas} (see also Fig.~\ref{diagramXAS}).  In the context of RIXS, we refer to this term as the exciton propagator dressed by virtual (intermediate-state) phonons.  This term includes the contributions of virtual phonons to infinite order.

When the final-state contains a non-zero number of phonons, the RIXS amplitudes may be written down following usual diagrammatic rules, supplying $M^{\nu^{\prime}}_{\xi\xi^{\prime}}$ for a vertex, $\Lambda_{\xi}$ for an exciton line and $D_{\nu^{\prime}}^{>}$ for a phonon propagator.
The term with one real (final-state) phonon (Fig.~\ref{norecoil_d1}b) is
\begin{equation}
\label{Y1}
\Lambda^{(1)}(\xi,\xi^{\prime},\nu^{\prime},t)=iM^{\nu^{\prime}}_{\xi\xi^{\prime}}\Lambda_{\xi}(t-\tilde{\tau})D^{>}_{\nu^{\prime}}(t-\tilde{\tau})\Lambda_{\xi^{\prime}}(\tilde{\tau}) 
\end{equation}
where internal time integration is implied by the shorthand notation $P(\tilde{\tau})=\int_0^t d\tau P(\tau)$ for a given propagator $P$.  The phonon Green's function $iD^{>}_{\nu'}(t-\tau)=\bra{0}Tb_{\nu'}(t)b^+_{\nu'}(\tau)\ket{0}$ is a causal one (half of the full phonon Green's function) and  corresponds to the propagation of a real phonon which is created at time $\tau$ and destroyed at time $t>\tau$.  For two real phonons (Fig.~\ref{norecoil_d1}c), we have
\begin{eqnarray}
\label{Y2}
\Lambda^{(2)}(\xi,\xi^{\prime},&\nu^{\prime}&,\nu^{\prime\prime},t) = i^2\sum _{\xi_1} M^{\nu^{\prime\prime}}_{\xi\xi_1} \Lambda_{\xi}(t-\tilde{\tau}_2) D^{>}_{\nu^{\prime\prime}}(t-\tilde{\tau}_2)  \nonumber \\
&\times& \Lambda_{\xi_1}(\tilde{\tau}_2-\tilde{\tau}_1) M^{\nu'}_{\xi_1\xi^{\prime}} D^{>}_{\nu^{\prime}}(t-\tilde{\tau}_1) \Lambda_{\xi^{\prime}}(\tilde{\tau}_1) \, .
\end{eqnarray}
One can build the higher terms $\Lambda^{(n)}$ contributing any number of real phonons to the RIXS final state by analogy.

The numerical evaluation of these terms for a periodic crystal is feasible, though cumbersome.  For a simpler demonstration, we take the no recoil limit, which physically involves zero momentum transfer scattering between excitons and Einstein type vibrational modes.  In the next subsection, we derive a more compact and tractable representation of the RIXS amplitudes for the no recoil approximation that also allows us to include some vertex corrections.  In the subsection after that we present numerical results for acetone and a general analysis of the weak coupling limit.

\subsection{No recoil approximation}
\label{sec:norecoil}

\begin{figure}[htb]\centering
\center{\includegraphics[width=1\linewidth]{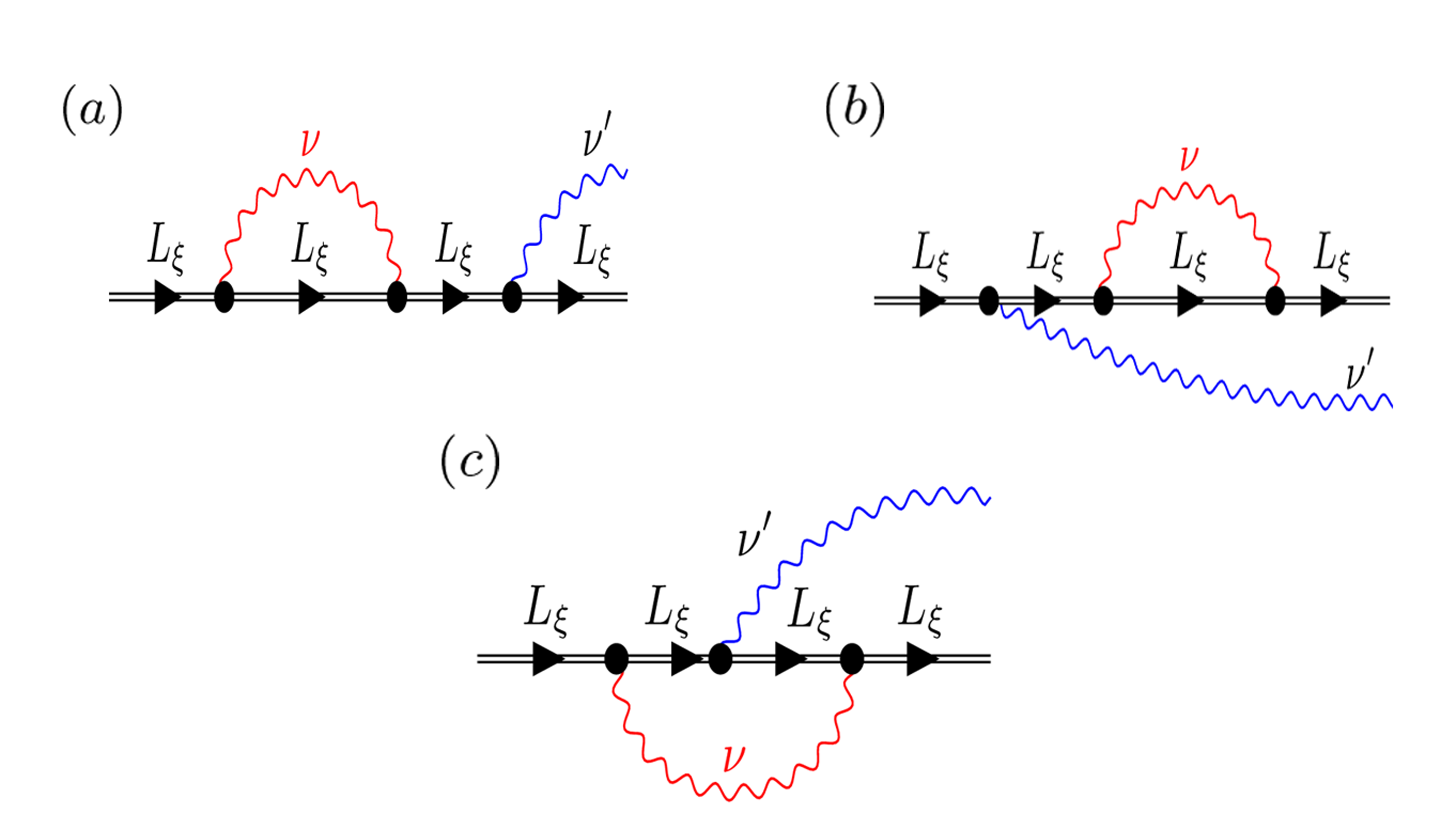}}	
\caption{Low order diagrams for exciton-phonon interactions during the RIXS process.  Black lines are bare exciton propagators, and red oscillating curves represent virtual phonons while blue oscillating curves are used for real phonons.  Diagrams $(a)$ and $(b)$ are contained within the diagram of Fig.~\ref{norecoil_d1}b for $\Lambda^{(1)}$ while $(c)$ is a vertex correction accounted for by Eq.~\eqref{vertex0}.}
\label{norecoil_d2}
\end{figure}

To clarify the following discussion, we draw a few low order contributions to $\Lambda^{(1)}$ in Fig.~\ref{norecoil_d2}, where we switch from the phonon-dressed exciton propagator $\Lambda_{\xi}$ to the bare exciton propagator $L_{\xi}$ and explicitly draw the virtual phonons in red (real phonons remain in blue).  The above construction of the RIXS amplitudes accounts exactly for diagrams such as those shown in Fig.~\ref{norecoil_d2}a and Fig.~\ref{norecoil_d2}b, as well as all replicas.  However, it neglects contributions from vertices where the emission of a real phonon occurs concomitantly with the excitation of a virtual phonon (Fig.~\ref{norecoil_d2}c).  

To account for these vertex corrections, we recognize that in the absence of recoil, $\Lambda^{(0)}$ takes the same form as the XPS core-hole Green's function.  Namely, $\Lambda^{(0)}_{\xi}(t) = L_{\xi}(t) \exp [C_{\xi}(t)]$ where the cumulant is given by Eq.~\eqref{norecoilcumulant} except for the substitution of exciton-phonon coupling constants for the core-hole--phonon coupling parameters.  Through a manipulation of the S-matrix expansion presented in Appendix C, the one real phonon contribution is
\begin{equation}
\label{lambda-1}
 \Lambda^{(1)}(\xi,\nu',t)=\Lambda^{(0)}(\xi,t){Y}(\xi,\nu',t) 
\end{equation}
where we have defined the vertex part
 \begin{equation}
\label{vertex0}
Y(\xi,\nu',t)=iM^{\nu'}_{\xi}\int_0^t D^{>}_{\nu'}(t-\tau)d\tau \ ,
\end{equation}
which involves only real (final-state) phonons.  Extending this result to an arbitrary number of final-state phonons the RIXS amplitudes can be expressed as
\begin{equation}
\label{lambda-n}
\Lambda^{(n)}(\xi,\nu',t)= \Lambda^{(0)}(\xi,t)\frac{[Y(\xi,\nu',t)]^n}{\sqrt{n!}} \ .
\end{equation}
This expression contains no explicit summation over virtual phonons, which is typically the most expensive part of RIXS calculations.  These effects are implicitly accounted for within $\Lambda^{(0)}$.  Contributions due to a particular number of final-state phonons may be evaluated separately up to arbitrary $n$.

\subsection{Numerical results}
\label{rixs:numerics}

\begin{figure}
\center{\includegraphics[width=1\linewidth]{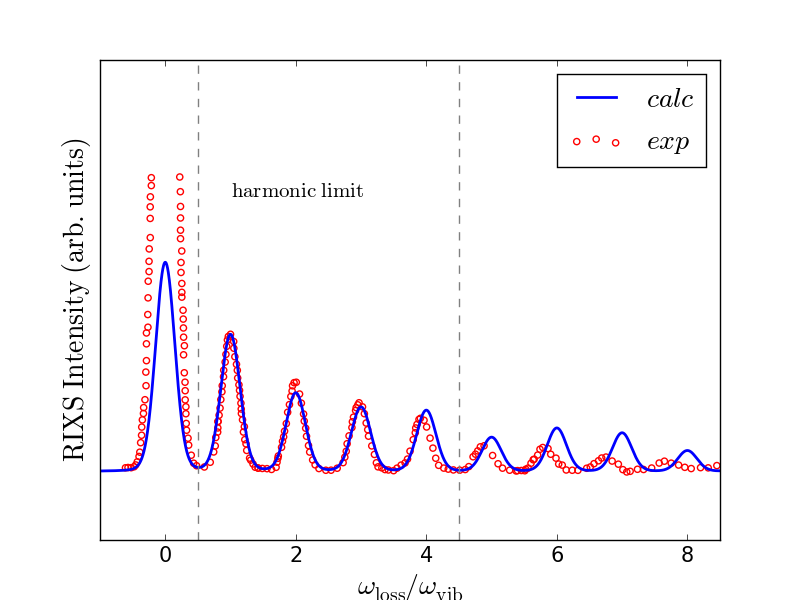}}

\caption{Calculated phonon contribution to the O-K edge RIXS spectrum of acetone (blue curve) compared to experimental results \cite{Schreck2016} (red symbols). Spectra were normalized to the intensity of the first phonon peak. Anharmonic effects become significant in the experimental data after the fourth phonon peak.}
\label{rixsgf}
\end{figure}

For numerical demonstration, we again chose the O-K edge of the acetone molecule, which involves the coupling of a single localized exciton to the C=O bond stretching mode.  Evaluation of Eq.~\eqref{rixsvib} requires the photon matrix elements and pure electronic excitation spectrum, exciton-phonon coupling constant and the phonon Green's functions for the electronic ground-state and the core-excited intermediate-state. These quantities were already obtained in Sec.~\ref{xas} during the calculation of the vibrational contribution to the O-K XAS spectrum of acetone.  Calculation of the XAS coefficient gives essentially $\Lambda^{(0)}$, to which we add a core-hole lifetime for the RIXS intermediate state.  The vertex part is then calculated according to Eq.~\eqref{vertex0} and the two quantities are combined in Eq.~\eqref{lambda-n} to give $\Lambda^{(n)}$.  Additional Gaussian broadening was added to the final spectrum, given in Fig.~\ref{rixsgf}, to account for the experimental resolution.  While the overall agreement with experiment is favorable, small differences in the intensities and peak positions at higher oscillator numbers can be attributed to the neglect of anharmonic contributions \cite{Schreck2016}.

For phonon modes in periodic crystals, it is reasonable to assume a single vibrational frequency for both the ground and core-excited states.  This will be less accurate for local vibrational modes in molecules.  However, evaluating the Franck-Condon overlap integrals for a model of acetone, we found that the differing curvatures of the ground and core-excited PES should make only minor corrections to the shape of the RIXS spectrum for the present case.  Consequently, we used the same frequency for both the real and virtual vibrations in the calculation for acetone.

\begin{figure}
\center{\includegraphics[width=1\linewidth]{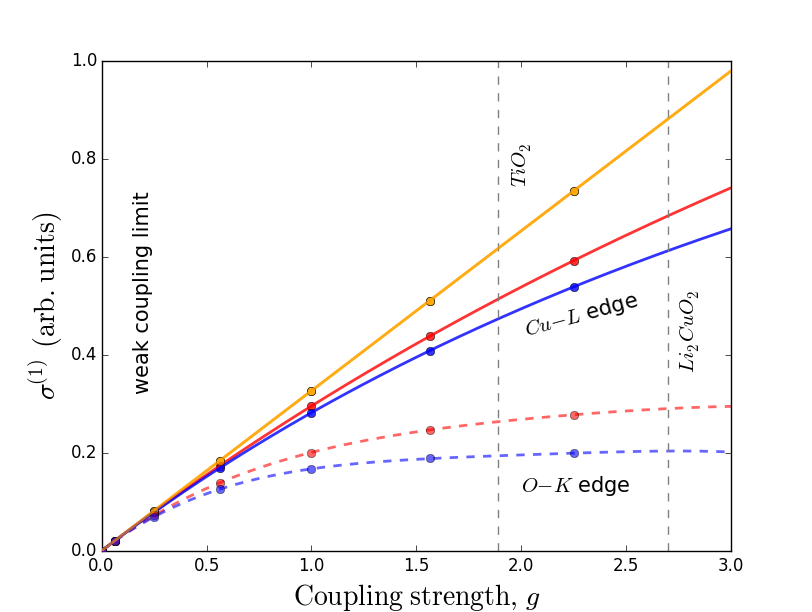}}	
\caption{Intensity of the first phonon peak with respect to the coupling strength.  Calculations were done for a single phonon mode in the no-recoil limit using different approximations: yellow line - lowest order contribution (no virtual phonons); red curves -  virtual phonons, but no vertex corrections (Eq.~\eqref{Y1}); blue curves - virtual phonons including vertex corrections (Eq.~\eqref{lambda-1}).  Dashed and solid lines correspond to different core-hole lifetimes.  The ratio $\Gamma_M/\omega_{vib}$ was set to 2 (dashed lines) and 6 (solid lines) as typical values for O-K and Cu-L edges, respectively.  All curves were consistently  normalized to the value of the lowest order contribution (yellow line).  Vertical dashed lines indicate values of the coupling strength measured by RIXS for TiO$_2$ \cite{Moser2015a} and Li$_2$CuO$_2$ \cite{Johnston2016a}.}
\label{firstpeak}
\end{figure}

In the limit of weak coupling, one can consider only the lowest order diagrams.  For example, Devereaux $et \ al.$ \cite{Devereaux2016} recently calculated the RIXS cross-section for a small cluster model of CuO in the limit of no virtual phonons and one final-state phonon.  In an effort to quantify the region of applicability of this weak coupling limit, we compare model calculations of the 1-phonon contribution to RIXS $\sigma^{(1)}$ as a function of coupling strength {\it i}) when all virtual phonons are neglected, {\it ii}) when virtual phonons are included to infinite order without vertex corrections, and {\it iii}) when virtual phonons are included to infinite order with vertex corrections.  These results are presented in Fig.~\ref{firstpeak}.  The yellow line gives the intensity of the first phonon peak when virtual phonons are neglected.  The red and blue lines correspond to the inclusion of all virtual phonons, without or with vertex corrections, respectively.  The solid lines are evaluated using a ratio $\Gamma_{M}/\omega_{ph}$ of the intermediate-state lifetime to phonon frequency appropriate for the Cu-L edge of a 2D cuprate while the dashed lines use a value consistent with the O-K edge.

We confirm that the weak coupling approximation is reasonable for coupling strengths less than about 1, however this depends on the ratio of the core-hole lifetime and the phonon frequency.  Due to the longer core-hole lifetime of the O 1s level, the curves for the O-K edge including virtual phonons deviate from the weak coupling approximation earlier than those for the Cu-L edge.  For the O-K edge, the values of $\sigma^{(1)}$ including virtual phonons already differ by a factor of 2 from the weak coupling approximation by $g=1$.  The deviation of the results obtained with all virtual phonons (red and blue curves) from the zero virtual phonon values (yellow line) indicates that the calculated $\sigma^{(1)}$ contribution is over-estimated since the 2-phonon contribution ($\sigma^{(2)}$) becomes non-negligible and takes spectral weight from $\sigma^{(1)}$.  If the coupling strength varies throughout the Brillouin zone the correction to the one phonon intensity will also vary in momentum space, likely making it important to go beyond the weak coupling limit.

For periodic systems with several active phonon modes at different frequencies it can be difficult to distinguish the second harmonic of low frequency modes from the first harmonic of higher frequency modes, making it non-obvious to experimentally identify weak coupling cases.  This partly explains why quantitative experimental studies of electron-phonon coupling by RIXS are still limited.  However, we note that some of these measurements report intermediate or even strong coupling values \cite{Lanzara2001, Lee2013, Moser2015a, Johnston2016a}.  For reference, in Fig.~\ref{firstpeak} we use vertical dashed lines to indicate experimentally obtained coupling parameters for TiO$_2$ and the quasi-1D Li$_2$CuO$_2$.

\section{Conclusions}
\label{conclusion}

We have studied the phonon contribution to the RIXS loss spectrum from the qualitative, quantitative and formal perspectives.  Contrary to common assertion, we find that RIXS is not a direct probe of electron-phonon coupling, even when measured at the Cu $L_3$ edge.  Both excitonic binding effects and direct core-hole--phonon interactions cause considerable deviations from the electron-phonon interpretation.  We find, however, that an exciton-phonon coupling description is able to quantitatively explain previous experimental data for acetone.  We expect that the exciton-phonon description of the RIXS interaction will hold in general.  This result significantly impacts the use of RIXS to quantify electron-phonon coupling strengths in cuprates and other materials.

The notion that RIXS intensities are reflective of electron-phonon coupling strengths in cuprates assumes that a Cu $2p_{3/2}$ core-hole, at about 930 eV binding energy, should be fully screened.  Experimental X-ray photoemission spectra \citep{Thomas2002} at the much shallower Si $2p$ level (around 100 eV) of silicon tetra-halides have shown obvious phonon sidebands, clearly signaling the ability of a core-hole to excite vibrations.  Our test calculations on Cu$_2$O, presented in Appendix B, suggest the contribution of the deeper Cu $2p$ hole to the electron-lattice interaction is comparable to that of the excited electron and cannot be neglected.  Although we find that the electron-lattice interaction probed by RIXS is better characterized as exciton-phonon coupling than electron-phonon coupling, we suspect that it will be possible to disentangle the latter quantity, which is the principle interest, from the former.

Based on the quantitative agreement of the calculated phonon contribution to RIXS -- assuming exciton-phonon coupling -- with experimental data for acetone, we have developed a many-body Green's function description of the phonon contribution to the RIXS cross section.  We employed a cumulant expansion for the exciton Green's function, in conjunction with a Fan-Migdal type exciton self-energy.  This methodology succeeded in accurately reproducing phonon satellite structure observed experimentally in the X-ray absorption spectrum of acetone, as well as the phonon excitation series measured by RIXS.  The Green's function formulation is advantageous compared to wavefunction-based calculations that require onerous summations of all possible RIXS intermediate states.  Our methodology includes only an explicit summation over the RIXS final states, which are limited to the lowest few phonon oscillator levels in practice.

We have demonstrated our formalism on acetone due to the availability of high quality and unambiguous experimental data.  However, we intend that the methodology be applied to periodic crystals.  Periodic systems present additional numerical challenges associated with sampling phonon coupling strengths throughout the Brillouin zone.  While computationally demanding, such sampling is still possible within the framework of first-principles calculations \cite{Nery2017}.

\section*{acknowledgments}

We thank Tim Ziman, Patrick Bruno and Nick Brookes for valuable discussions and close reading of the manuscript.

\section*{appendix A: Numerical Details}

%\subsection*{Numerical Details}

All density functional theory, density functional perturbation theory, and {\it ab initio} molecular dynamics calculations were performed with the Quantum-ESPRESSO package \cite{Giannozzi2009}, which employs pseudopotentials, a planewave basis and periodic boundary conditions.  Unless stated otherwise, we used ultrasoft, PBE/GGA pseudopotentials taken from the Quantum-ESPRESSO pseudopotential library, and a planewave cutoff of 50 Ry for the wavefunctions and 400 Ry for the charge density.

To evaluate the excited-state potential energy surfaces and force constants in Sec.~\ref{sec:intro}, a single acetone molecule was placed in a cubic box of length 20 ${\rm \AA}$ and atomic positions were relaxed to their ground-state equilibrium positions.  We used the frozen-phonon scheme in which the oxygen atom was given several displacements from its equilibrium position and the total energy evaluated at each displacement for each of the three excited-state scenarios (extra electron, core-hole and exciton).  The resulting forces are presented in Table \ref{ForceTable}.  The core-hole and exciton cases required the use of an oxygen pseudopotential with one $1s$ core electron removed.  The extra electron and exciton configurations constrained an additional electron to the LUMO level, which we confirmed had significant overlap with an anti-bonding C-O molecular orbital in both cases.
 
To preform the Born-Oppenheimer MD simulations for the XPS spectral functions in Sec.~\ref{XPSsect} we constructed an ultrasoft, PBE Si pseudopotential with a $2p$ hole.  Calculations were performed to mimic the gas-phase by placing a single molecule in the center of a 15 ${\rm \AA}$ vacuum cell.  We used an MD time step of 50 $\hbar / E_{\rm Ryd}$.  The bare core-hole spectral function used in the convolution with the vibronic spectral function was constructed by hand from experimental knowledge of the Si $2p$ spin-orbit splitting and binding energies.  Values for the spin-orbit splitting and the respective binding energies (which depend moderately on the local chemical environment) could be obtained numerically, $e.g.~$from all-electron DFT or Hartree-Fock calculations.

To obtain the exciton-phonon force constants used in Sec.~\ref{xas} and Sec.~\ref{RIXS}, Bethe-Salpeter calculations for the X-ray absorption spectra of N$_2$, CO and acetone were performed using the OCEAN code \cite{Shirley1998, Vinson2011, Gilmore2015} with Quantum-ESPRESSO as the underlying DFT engine.  Both molecules were treated as gas-phase by placing them in a cubic supercell of side length 20 ${\rm \AA}$.  Ground-state (non-core-hole) LDA, norm-conserving pseudopotentials were used with an energy cutoff of 100 Ry.  Convergence was reached for acetone by including 96 unoccupied bands for the core-hole screening calculation and 72 bands for the exciton basis.

In certain cases, such as those involving a core-hole in a $1s$ level, an alternative approach to obtaining the exciton-phonon coupling constants is to generate the excited-state potential energy surface with simpler self-consistent DFT calculations using a core-hole pseudopotential and placing an extra electron in the LUMO orbital \cite{Prendergast2006, Gougoussis2009}.  We extracted the force constants from the excited state potential energy surfaces using both the BSE and constrained core-hole--excited-electron approaches.  The exciton forces may be also estimated from the Hellmann-Feynman theorem in the core-hole--excited-electron case and then project onto normal vibrational coordinates following the procedure described in Sec.~\ref{sec:intro}.
\begin{table}
\caption{\label{tab:forces}Vibronic forces for acetone obtained from the slope of the excited-state PES (CO stretching mode) evaluated at the ground-state equilibrium bond length with respect to different types of excitations (see Sec.~\ref{sec:intro}).}
\begin{ruledtabular}
\begin{tabular}{ll}
\label{ForceTable}
\textrm{Excitation type}&
\textrm{Force $\ {\rm (eV/\AA)}$}\\
\colrule
electron & -1.0\\
core-hole & -3.5\\
exciton & -7.7\\
\end{tabular}
\end{ruledtabular}
\end{table}

\section*{Appendix B: ${\rm Cu}$ $2p$ core-hole in ${\rm Cu_2O}$}

Our study of the O-K edge RIXS of acetone clearly shows the importance of direct coupling of the core-hole to vibrations. Since many RIXS studies are performed at the Cu $L_3$ edge, we now estimate the contribution from a Cu $2p$ hole in generating lattice dynamics.  We select crystalline Cu$_2$O for this test and approximate the forces on nearest neighbor oxygen sites due to an excitation on a copper site.  As in the acetone example from Sec.~\ref{sec:intro}, we model an extra electron, a core-hole, and an exciton.  The resulting forces are given in Table \ref{tab:cu}.  The force resulting from the introduction of a Cu 2$p$ core-hole is approximately 60 $\%$ of the force due to an exciton, and slightly larger than the force from the addition of an electron localized at the copper site.  From this, we conclude that the Cu $2p$ hole is not screened enough to neglect it's coupling to phonons.  The details of these calculations are provided in the following paragraphs.

\begin{table}[h]
\caption{\label{tab:cu}Forces on the nearest neighbor oxygens due to an electronic excitation on a Cu site in crystalline ${\rm Cu_2O}$. }
\begin{ruledtabular}
\begin{tabular}{ll}
\textrm{Excitation type}&
\textrm{Force $\ {\rm (eV/\AA)}$}\\
\colrule
effective electron & 0.135\\
core-hole & 0.165\\
exciton & 0.257\\
\end{tabular}
\end{ruledtabular}
\end{table}
DFT calculations were performed on a ($3\times3\times3$) supercell of ${\rm Cu_2O}$ using LDA norm-conserving pseudo-potentials.  The experimental structure was initially relaxed in order to minimize the electron-ion forces for the ground-state configuration.  Keeping the ground-state atomic positions fixed, SCF calculations were made for core-hole and 'exciton' configurations.  For these, we made a copper pseudopotential with one electron removed from the $2p$ shell and used this pseudopotential on one of the 108 copper sites in the supercell.  The core-hole configuration has an overall positive charge of $1|e|$ for the supercell that can be compensated by a uniform negative charge.  Explicitly adding an extra electron to the bottom of the conduction band (instead of using a uniform neutralizing charge density) gives the exciton configuration since this extra electron will be localized around the core-hole site.  For each configuration, we obtain the force on the oxygen atoms nearest to the copper site with the core-hole.  We find that the force on the nearest oxygen atoms for the core-hole configuration is 64 $\%$ of the force resulting from the exciton configuration.  This strongly suggests that even for the deeply bound Cu $2p$ levels a core-hole is not completely screened and contributions directly to the generation of phonons.

To mimic the addition of an extra, localized electron, to otherwise neutral Cu$_2$O, we substituted Zn for one of the Cu sites. Due to the impurity nature of the Zn atom, the highest occupied electron level is localized around the zinc site.  We repeated the calculation after removing this electron (giving a supercell with total positive charge of $1|e|$) and defined the force due to the excited electron as the difference between forces on nearest oxygen atoms for the neutral and charged (Zn and Zn$^{+1}$) impurity configurations ($F_e=(F_{Zn-O})-(F_{Zn^{+1}-O})$).  The resulting force on the nearest oxygen site is smaller than for either the exciton or core-hole configurations, possibly due to the less localized nature of the extra electron compared to the bound exciton.  Although more effective schemes for localizing an extra electron can likely be constructed with a local orbital basis DFT code, the comparison between the core-hole and exciton configurations already indicates that the core-hole--phonon coupling should not be neglected at the Cu $L_3$ edge.

\section*{Appendix C: Zero momentum transfer RIXS}
\label{non-diagonal-c}
In order to clarify the origin of Eq.~\eqref{lambda-1}, we  recall the S-matrix expansion and consider the $n$=0 and $n$=1 elements of the $\Lambda^{(n)}$ series.  In the zero momentum transfer limit, the interaction part of the Hamiltonian is $V=\sum_{\nu,\xi} M^{\nu}_{\xi\xi} a_\xi^+a_\xi B_\nu$.  The term with zero real phonons may be written formally as
\begin{widetext}
\begin{equation}
\Lambda^{(0)}(\xi,\xi,t)= -i\sum^{\infty}_{m=0}
 \frac{(-i)^{2m}}{(2m)!} \int_0^t dt_1 \dots \int_0^t dt_{2m} 
  \bra{0}T a_\xi(t)V(t_{2m}) \cdots V(t_1)a^+_\xi(0)\ket{0}  \ .
\label{lambda}
\end{equation}
Since $H^0$ commutes with the number operator $a^+_\xi a_\xi$ and all exciton lines have the same index $\xi$, we may factor out the bare exciton Green's function $iL_\xi(t)=\bra{0}Ta_\xi(t)a^+_\xi(0)\ket{0}$ from this series \cite{Nozieres1969,Langreth1970,Mahan1990} leaving
\begin{equation}
\Lambda^{(0)}(\xi,\xi,t)=L_\xi(t)\sum^{\infty}_{m=0}
\frac{(-i)^{2m} M^{\nu_1}_{\xi\xi} \cdots M_{\xi\xi}^{\nu_{2m}}}{(2m)!}  \sum_{\nu_1 \dots \nu_{2m}} \int_0^t dt_1  \cdots \int_0^t dt_{2m} 
\bra{0}TB_{\nu_{2m}}(t_{2m}) \cdots B_{\nu_1}(t_1)\ket{0} \ .
\label{lambda-xi-n=0}
\end{equation}
After pairing all phonon operators into phonon Green's function, the series may be summed up, giving an exponential generating function $\exp[C(t)]$ which involves only interactions with virtual phonons \cite{Langreth1970}.  Thus, $\Lambda^{(0)}(\xi,\xi,t)=L_\xi(t) e^{C(t)}$. 
For the one real phonon case ($n$=1), two types of phonon Green's function will be present in the expansion of Eq.~\ref{lambda} after pairing all phonon operators. One is related to the propagation of virtual phonons and other describes the real phonon. Using the same argument as before, we factor the bare exciton propagator out of the series 
\begin{eqnarray}
\Lambda^{(1)}(\xi,\xi,\nu^{\prime},t)=L_\xi(t)\sum^{\infty}_{m=0}
 \frac{(-i)^{2m} M^{\nu_1}_{\xi\xi} \cdots M_{\xi\xi}^{\nu_{2m}}}{(2m+1)!} \sum_{\nu_1 \dots \nu_{2m+1}}
  (&2m&+1)\int_0^t dt_{2m+1} M^{\nu'}_\xi \bra{0}TB_{\nu_{2 m+1}}(t_{2m+1})b_{\nu'}(t)\ket{0}\delta_{\nu'\nu_{2m+1}}  \nonumber \\ 
 &\times& \int_0^t dt_1 \dots \int_0^t dt_{2m} 
 \bra{0}TB_{\nu_{2m}}(t_{2m}) \cdots B_{\nu_1}(t_1)\ket{0} \ .
\label{lambda-xi-n=1}
\end{eqnarray}
\end{widetext}
The main difference between $\Lambda^{(1)}$ and $\Lambda^{(0)}$ is the presence of the first integral, with integration variable $t_{2m+1}$.  This integral contains the real phonon with one vertex and a single time integration variable, and may be condensed to the expression 
\begin{equation}
\label{vertex:appendix}
Y(\xi,\nu',t)=iM^{\nu'}_{\xi}\int_0^t D^{>}_{\nu'}(t-\tau)d\tau \ .
\end{equation}
Since this $Y$ factor does not involve virtual phonons it may be considered separately.  It appears 2$m$+1 times in the (2$m$+1)-th term in the expansion of Eq.~\eqref{lambda-xi-n=1} and combines with $\frac{1}{(2m+1)!}$ to give a final factor of $\frac{1}{(2m)!}$.  If we factor this term out of the sum the remaining terms are the same as the $\Lambda^{(0)}$ contribution for the virtual phonon dressed exciton propagator.  This gives
\begin{equation}
\label{lambda-1:appendix}
\Lambda^{(1)}(\xi,\xi,\nu',t)=\Lambda^{(0)}(\xi,\xi,t) Y(\xi,\nu',t) \ .
\end{equation}
Extending this result to an arbitrary number of final state phonons we arrive at Eq.~\eqref{lambda-n}
\begin{equation}
\label{lambda-n:appendix}
\Lambda^{(n)}(\xi,\xi,\nu',t)=\Lambda^{(0)}(\xi,\xi,t) \frac{[Y(\xi,\nu',t)]^n}{\sqrt{n!}} \ .
\end{equation}
The factor $\frac{1}{\sqrt{n!}}$ comes from the normalization of the final-state vibrational wave-function.

%\bibliographystyle{aipnum}

%\bibliography{full_4}
\bibliography{ref_full}
%\bibliography{hed,cumgw,srvo,newup,raman4,ref_full}

\end{document}